\documentclass[twocolumn]{aa}  

\newcommand{\kms}{\ensuremath{\mathrm{km\,s^{-1}}}}
\usepackage{graphicx}
\usepackage{txfonts}
\usepackage{lipsum}
\usepackage{subcaption}         
                                
\usepackage{lscape}            
                                
\usepackage{placeins}          
                                
\usepackage{natbib,twoopt}   
\bibpunct{(}{)}{;}{a}{}{,}
\usepackage[colorlinks, citecolor={blue}]{hyperref}
\usepackage{etoolbox}
\AtBeginDocument{\nolinenumbers}
\apptocmd{\maketitle}{\nolinenumbers\let\linenumbers\relax}{}{}

\begin{document}

   \title{Exploring the formation mechanisms of tidal structures in globular clusters of extragalactic origin}

   \author{Shouzhi Wang\inst{1,2,3}
        \and Jundan Nie\inst{3}\fnmsep\thanks{jdnie@nao.cas.cn} \and Biwei Jiang\inst{1,2} \and Hao Tian\inst{3} \and Chao Liu\inst{3} \and Ying-Hua Zhang\inst{3,4}}

   \institute{Institute for Frontiers in Astronomy and Astrophysics, Beijing Normal University, Beijing 102206, People's Republic of China
   \and School of Physics and Astronomy, Beijing Normal University, Beijing 100875, People's Republic of China
   \and National Astronomical Observatories, Chinese Academy of Sciences, Beijing 100101, People's Republic of China \\ \email{jdnie@nao.cas.cn}
   \and School of Astronomy and Space Science, University of Chinese Academy of Sciences, Beijing 100048, People's Republic of China}

   \date{}

  \abstract
{Tidal structures around globular clusters provide valuable insights into cluster evolution and the hierarchical assembly of the Milky Way. Using wide-field imaging data from the DESI Legacy Survey combined with a color-magnitude matched-filter technique, we performed a systematic analysis of extratidal features in 28 Galactic globular clusters of likely extragalactic origin, representing the largest homogeneous sample examined in this context to date. The clusters display diverse morphologies: 12 exhibit tidal tails, nine show diffuse envelopes, and seven reveal no clear extratidal features. Notably, we report the first detection of an extended tidal structure around the Sagittarius-associated cluster Terzan 7. To explore the underlying drivers, we compared intrinsic properties, orbital dynamics, and possible accretion associations across morphological groups. From the parameter distributions, complemented by Kolmogorov-Smirnov tests, we find that total mass, escape velocity, concentration, tidal filling factor, pericentric radius, eccentricity, and radial angle in action-angle coordinates are all likely group-sensitive parameters. These results suggest that both internal cluster properties and orbital configurations play important roles in shaping extratidal morphologies. In addition, the cluster's accretion history shows no clear correlation with the presence of tidal features, indicating that it is not a direct driver of outer structure formation. Overall, the diversity of tidal structures is unlikely to be governed by a single factor. Instead, it reflects the interplay between internal dynamical evolution and the external Galactic environment. This study provides the most comprehensive constraints so far on the physical processes driving extratidal structures in accreted globular clusters.}

   \authorrunning{Wang et al. }            
   \titlerunning{Extratidal structures of globular clusters}           
   \maketitle

\section{Introduction}
Globular clusters (GCs) are among the oldest stellar systems in the Universe, and they serve as valuable tracers of the formation and assembly history of their host galaxies. In the hierarchical framework of Lambda cold dark matter cosmology model \citep{2016A&A...594A...1P}, many Galactic GCs are thought to have been accreted from satellite galaxies during past merger events \citep{1978ApJ...225..357S, 2005ApJ...635..931B, 2020MNRAS.498.2472K}. This idea is supported by the discovery of phase-space substructures and by the association of certain clusters with known merger remnants, such as the Sagittarius dwarf galaxy or Gaia-Sausage-Enceladus \citep[see, e.g.,][and references therein]{2010ApJ...718.1128L, 2018ApJ...862...52S, 2018Natur.563...85H, 2019A&A...630L...4M}.
These GCs of extragalactic origin
are believed to have complex merger histories and dynamical perturbations, which may render them more susceptible
to developing prominent extratidal features under the influence
of tidal forces \citep{1978ApJ...225..357S, 2016MNRAS.463.1759B, 2017ApJ...847..119G, 2017MNRAS.465.3622R, 2018ApJ...861...69C}. 

Tidal structures are generally believed to form when stars acquire sufficient energy to escape the gravitational potential of their host GCs. In the early stages of cluster evolution, energetic feedback from stellar evolution, gas expulsion, and dynamical disequilibrium can cause young clusters to rapidly lose their gravitational binding and dissolve \citep{2001MNRAS.323..988G, 2001MNRAS.322..231K, 2006MNRAS.373..752G, 2007MNRAS.380.1589B}. Over longer timescales, stellar escape can be driven by a combination of internal and external mechanisms. Internally, two-body relaxation redistributes energy among stars, enabling some to reach escape velocity \citep{2008MNRAS.389..889K, 2008MNRAS.387.1248K}. Externally, tidal stripping induced by the Galactic potential as well as dynamical perturbations such as disk and bulge shocking during the cluster's orbital passages through the Galactic plane or central regions can further enhance stellar mass loss and facilitate the development of extratidal features \citep{1997MNRAS.289..898V, 2003MNRAS.340..227B, 2007ApJ...659.1212M, 2015MNRAS.446.3100H}. 

Due to the complex formation pathways and the diverse Galactic environments in which they reside, GCs exhibit a wide variety of tidal structures, such as symmetric or multi-armed tidal tails and diffuse outer envelopes. However, the physical mechanisms driving the formation of these different tidal structures are still under active investigation. Early numerical N-body simulations by \citet{2007ApJ...659.1212M} suggest that repeated pericentric passages or strong interactions with the dense regions of the Galaxy can trigger the stripping of outer stars, leading to the formation of tidal tails. In contrast, the work by \citet{2010MNRAS.401..105K, 2012MNRAS.420.2700K} highlights the role of epicyclic motions of escaping stars in shaping the formation and morphology of tidal features. Other studies have pointed out that orbital parameters, such as inclination and eccentricity, play an important role in regulating the mass loss of clusters \citep{2019ApJ...882...98P, 2020A&A...637L...2P}, which in turn affects their susceptibility to developing tidal structures. Meanwhile, the potential connection between cluster origin and structural properties has garnered increasing attention. For example, whether the presence of tidal structures is connected to specific accretion events, such as the Sagittarius dwarf galaxy or the Gaia-Sausage-Enceladus merger, remains to be systematically investigated.

Benefiting from the advent of wide-field imaging surveys such as the Sloan Digital Sky Survey (SDSS; \citealt{2009ApJS..182..543A}), Pan-STARRS \citep{2016arXiv161205560C}, the Dark Energy Survey (DES; \citealt{2018ApJS..239...18A}), and the DESI Legacy Imaging Surveys \citep{2019AJ....157..168D}, together with the development of various techniques optimized for identifying tidal substructures, most notably the application of matched filter method, previous studies have uncovered an increasingly rich diversity of tidal features, including long tidal tails (e.g., Palomar 5 and NGC 5466; \citealt{2006ApJ...641L..37G,2006ApJ...639L..17G}), multi-arm structures (e.g., NGC 4147; \citealt{2010A&A...522A..71J, 2024AJ....168..237Z}), and a variety of other complex substructures reported in the literature \citep{2006ApJ...651L..33L,
2009AJ....138.1570O,
2010AJ....139..606C,
2010A&A...522A..71J,
2012MNRAS.419...14C,
2014MNRAS.445.2971C,
2016MNRAS.461.3639K,
2017MNRAS.467L..91C,
2017ApJ...841L..23N,
2018MNRAS.474..683C,
2018MNRAS.473.2881K,
2018ApJ...860...66M,
2018MNRAS.473.3062M,
2018ApJ...862..114S,
2019MNRAS.486.1667C,
2019MNRAS.485.4906D,
2019ApJ...884..174G,
2019MNRAS.488.1535P,
2020A&A...635A..93P,
2020AJ....160..244S,
2020MNRAS.495.2222S,
2020ApJ...902...89T,
2021ApJ...914..123I,
2021A&A...646A.176P,
2022ApJ...930...23N,
2022MNRAS.509.3709P,
2022MNRAS.513.3136Z,
2025A&A...693A..69A,
2025AJ....170..294C,
2025MNRAS.540.2863W,
2025ApJ...988...39W,
2025ApJ...984..189Y}. Precisely because such a large sample of GCs with diverse tidal morphologies has now become available, we are afforded the opportunity to investigate, from an observational perspective, the key physical mechanisms that drive the diversity of tidal structures in these systems.

Recent studies have utilized observational data to explore the key factors that drive the formation of different types of tidal structures. \citet{2020A&A...637L...2P} conducted a multi-parameter analysis of 53 Galactic GCs compiled from previous literature, including both in situ and accreted systems. However, their classification of tidal structures relied on results from multiple studies that used heterogeneous datasets obtained with different telescopes. More recently, \citet{2025AJ....170..157K} constructed a smaller sample of 30 GCs from the Pristine-Gaia-Synthetic catalog, likewise covering both in situ and accreted clusters, and in a brief section of their work, they also performed an exploratory parameter correlation analysis. These studies have provided valuable attempts to explore possible connections between cluster properties and tidal structures, although no significant correlations were found. Nevertheless, studies based on homogeneous data that focus exclusively on a specific class of globular cluster systems, such as those accreted from outside, are still lacking. Such clusters, owing to their external origin, may have undergone distinct pathways of dynamical evolution and tidal disruption \citep{2019A&A...630L...4M, 2020MNRAS.493..847F}.

The objective of this paper is to investigate the key factors that give rise to the tidal structures of the accreted globular cluster population by employing a homogeneous dataset and a unified methodology. We used the deepest dataset among the currently available public data and the largest accreted cluster sample among those widely recognized \citep{2019A&A...630L...4M, 2020MNRAS.493..847F}. Combined with a consistent matched-filter technique, this study minimizes the methodological biases and systematic uncertainties. 

This paper is organized as follows. In Sect. 2 we present the selection criteria for the globular cluster sample and introduces the photometric data employed. In Sect. 3 we outline the matched-filter technique adopted to search for tidal tails around GCs. In Sect. 4 we present the main results, including the identification and classification of tidal structures, their correlation with physical parameters, orbital properties, and the Galactic environment, as well as a discussion of the limitations in detecting such structures.
In Sect. 5 we summarize the main findings of this study.

\section{Observational material}\label{sec2}

We began by constructing an initial list of candidate GCs in the Galaxy with probable extragalactic origins based on consistent classifications in both \citeauthor{2019A&A...630L...4M} (\citeyear{2019A&A...630L...4M}, \citeyear{2025RNAAS...9...64M} update) and \citet{2020MNRAS.493..847F}. In the original 2019 version, \citet{2019A&A...630L...4M} classified clusters primarily according to their positions in integrals-of-motion (IOM) space derived from Gaia DR2 astrometry, with the age-metallicity relation serving as a secondary criterion. In the 2025 update, this classification was refined using Gaia eDR3 astrometry together with two independent sets of distance estimates from \citet{2021MNRAS.505.5957B} and \citeauthor{1996AJ....112.1487H} (\citeyear{1996AJ....112.1487H}, 2010 edition). We retained only those clusters that were consistently classified as accreted under both distance assumptions. In contrast, \citet{2020MNRAS.493..847F} adopted the IOM framework of \citet{2019A&A...630L...4M} but also incorporated the updated compilation of cluster ages and metallicities from \citet{2019MNRAS.486.3180K}, increasing the number of clusters with complete chemical information. Where available, they included $\alpha$-element abundances from \citet{2020MNRAS.493.3363H}, enabling more reliable assignments of clusters to disrupted dwarf galaxies and, in some cases, revising the original classifications. By cross-matching the results of these two studies and excluding clusters with unclear or ambiguous origins, we obtained an initial sample of 44 GCs of extragalactic origin.

For the search of tidal structures associated with these accreted GCs, we employed the DESI Legacy Imaging Surveys \citep{2019AJ....157..168D}, which represents the deepest photometric survey with the largest sky coverage currently available to the public. It combines three public optical surveys (DECaLS, BASS, and MzLS) to provide $g$-, $r$-, and $z$-band imaging over $\sim$14,000 deg$^2$ at Galactic latitudes $|b|>20^\circ$. The typical 5$\sigma$ point-source detection limits are $g=24.7$, $r=23.9$, and $z=23.0$ mag. Although the dataset is derived from multiple surveys, they were conducted and processed within the unified framework of the DESI Legacy Imaging Surveys, adopting a consistent observing strategy, depth and completeness requirements, data processing pipeline, and catalog construction procedures. These uniform procedures ensure that the combined photometric dataset has good homogeneity \citep{2019AJ....157..168D}. For this study, we primarily adopted DR8, which employs a source-detection-based photometric method that better preserves faint stars in crowded fields. In contrast, the Gaia-based forced photometry used in DR9 and DR10 tends to miss lower main-sequence stars, reducing both the photometric depth and completeness of the color-magnitude diagrams (CMDs) for several clusters (e.g., NGC 362, NGC 5897). In such cases, clusters available only in DR9 or DR10 were included in our sample only if their main sequence extends at least one magnitude below the turnoff, ensuring sufficient CMD depth and uniform photometric completeness.

We selected stellar sources classified as point-source in the DESI Legacy Surveys catalogs, requiring $g \leq 24$ and $r \leq 23$, with photometric uncertainties of less than 0.2 mag in both bands, in order to ensure a more complete and reliable star sample. For clusters lacking $r$-band data (e.g., Ter 7, Ter 8, and Pal 12), we substituted $i$-band photometry and applied the same selection criteria to maintain consistency. In the DESI Legacy Surveys catalogs, some photometric measurements (such as those in the $i$-band) are drawn from supplementary datasets incorporated into the survey products (e.g., the DES, the DELVE Survey, and the DeROSITA Survey). Subsequently, all magnitudes were corrected for Galactic extinction using the dust maps of \citet{1998ApJ...500..525S}, recalibrated by \citet{2011ApJ...737..103S}, with extinction coefficients of 3.214, 2.165, 1.592, and 1.211 for the $g$, $r$, $i$, and $z$ bands, respectively, following the calibrations of \citet{2019AJ....157..168D}.

Finally, we examined whether each cluster in the initial sample lies within the footprint of the DESI Legacy Imaging Surveys and possesses sufficient photometric coverage. Among them, 11 clusters (NGC 2298, NGC 3201, NGC 4833, NGC 5286, NGC 6235, NGC 6333, IC 1257, FSR 1758, ESO-SC06, NGC 6715, and NGC 6779) are located in regions not covered by DESI Legacy Surveys. In addition, NGC 2808 and NGC 6101 are covered only in the $g$ and $i$ bands, respectively; NGC 5139 suffers from severely incomplete coverage, with many areas lacking photometric detections; IC 4499 is heavily contaminated by field stars and has poor photometric quality; and NGC 2419, at a distance of about 88 kpc \citep{2021MNRAS.505.5957B}, is too remote for our photometric selection criteria to be reliably applied. After excluding these systems, we obtained a final working sample of 28 clusters (see Table \ref{tab1}). These include clusters with well-established progenitors (such as Gaia-Sausage-Enceladus (GSE), Sagittarius (Sag), Sequoia (Seq), and the Helmi streams (H99)) as well as clusters of clear accreted origin but uncertain association with a specific progenitor. Although some previous compilation studies have reported a comparable number of accreted GCs, the sample constructed in this work is entirely based on uniformly deep imaging data from the DESI Legacy Surveys, making it the deepest and largest homogeneous dataset to date for studying the tidal structures of GCs of extragalactic origin within a consistent data framework.

\begin{table*}
\caption{\raggedright Internal parameters and classification of our globular cluster sample.}
\label{tab1}
\centering
\resizebox{\linewidth}{!}{
\begin{tabular}{lcccccccccccc}
\hline\hline
Name  & $r_{\rm t}^{\rm King}$ & $r_{\rm t}^{\rm dyn}$ &$\log_{\rm 10}(M/M_\odot)$&$v_{\rm esc}$& $c$ &$r_{\rm h,m}/r_{\rm t}^{\rm dyn}$&$M_{\rm dis}/M_{\rm init}$& [Fe/H] &$\log_{\rm 10}({\rm age}/{T_{\rm rh}})$ &Type&Progenitor\\
      & ($'$) & ($'$) & & (\kms)&&&&(dex)&& \\
\hline
NGC 288   &   13.19  &  36.39 &  4.98 &  10.9 & 0.99    &0.09 &  0.26 &  -1.32 & 0.56 &G1&GSE    \\               
NGC 362   &   10.36  &  33.85 &  5.40 &  34.0 & 1.76    &0.04 &  0.28 &  -1.26 & 1.06 &G2&GSE    \\                
Whiting 1 &   0.89   &  4.94  &  3.13 &  1.1  & 0.55    &0.38 &  0.21 &  -0.70 & 0.48 &G1&Sag    \\               
NGC 1261  &   5.06   &  28.95 &  5.24 &  21.4 & 1.16    &0.04 &  0.19 &  -1.27 & 0.84 &G1&GSE    \\               
NGC 1851  &   6.52   &  35.58 &  5.45 &  41.1 & 1.86    &0.03 &  0.23 &  -1.18 & 1.06 &G2&GSE    \\                
NGC 1904  &   8.02   &  17.40 &  5.26 &  27.6 & 1.70    &0.07 &  0.44 &  -1.60 & 0.93 &G1&GSE    \\               
NGC 4147  &   6.08   &  17.38 &  4.65 &  14.2 & 1.83    &0.04 &  0.35 &  -1.80 & 1.48 &G1&GSE    \\                 
NGC 4590  &   14.91  &  25.88 &  5.11 &  14.8 & 1.41    &0.09 &  0.06 &  -2.23 & 0.58 &G2&H99    \\               
NGC 5024  &   18.37  &  34.01 &  5.70 &  25.9 & 1.72    &0.05 &  0.00 &  -2.10 & 0.16 &G3&H99    \\               
NGC 5053  &   11.43  &  17.96 &  4.80 &  6.0  & 0.74    &0.19 &  0.11 &  -2.27 & 0.20 &G2&H99    \\               
NGC 5272  &   28.72  &  43.65 &  5.61 &  32.0 & 1.89    &0.04 &  0.02 &  -1.50 & 0.57 &G3&H99    \\               
NGC 5466  &   15.68  &  16.36 &  4.75 &  6.5  & 1.04    &0.18 &  0.15 &  -1.98 & 0.35 &G1&Seq    \\               
NGC 5634  &   10.57  &  25.50 &  5.39 &  24.2 & 2.07    &0.04 &  0.09 &  -1.88 & 0.47 &G3&GSE/H99\\               
NGC 5897  &   10.14  &  19.73 &  5.22 &  12.5 & 0.86    &0.15 &  0.18 &  -1.90 & 0.32 &G1&GSE    \\               
NGC 5904  &   23.63  &  37.36 &  5.59 &  30.4 & 1.73    &0.07 &  0.05 &  -1.29 & 0.56 &G3&GSE/H99\\               
NGC 6205  &   21.01  &  60.36 &  5.69 &  32.4 & 1.53    &0.04 &  0.12 &  -1.53 & 0.61 &G3&GSE    \\               
NGC 6229  &   3.79   &  21.87 &  5.39 &  25.3 & 1.50    &0.03 &  0.20 &  -1.47 & 0.83 &G2&GSE    \\               
NGC 6341  &   12.44  &  45.70 &  5.44 &  36.4 & 1.68    &0.03 &  0.21 &  -2.31 & 1.00 &G2&GSE    \\                
Terzan 7  &   4.17   &  8.78  &  4.35 &  4.6  & 0.93    &0.20 &  0.08 &  -0.32 & 0.39 &G1&Sag    \\               
Arp 2     &   9.03   &  10.41 &  4.59 &  4.5  & 0.88    &0.22 &  0.08 &  -1.75 & 0.26 &G1&Sag    \\               
Terzan 8  &   3.98   &  13.51 &  4.88 &  6.0  & 0.60    &0.20 &  0.03 &  -2.16 & 0.01 &G1&Sag    \\              
NGC 6864  &   5.68   &  26.14 &  5.66 &  48.4 & 1.80    &0.02 &  0.14 &  -1.29 & 1.04 &G2&GSE    \\               
NGC 6981  &   7.46   &  11.14 &  4.91 &  12.4 & 1.21    &0.11 &  0.40 &  -1.42 & 0.97 &G2&GSE/H99\\               
NGC 7089  &   12.45  &  32.57 &  5.80 &  43.6 & 1.59    &0.04 &  0.16 &  -1.65 & 0.62 &G2&GSE    \\               
NGC 7099  &   18.97  &  28.93 &  5.08 &  21.0 & 2.50    &0.06 &  0.26 &  -2.27 & 1.10 &G3&GSE    \\               
Pal 12    &   19.10  &  11.89 &  3.79 &  2.5  & 2.98    &0.16 &  0.22 &  -0.85 & 0.82 &G3&Sag    \\               
Pal 13    &   2.19   &  5.51  &  3.44 &  1.8  & 0.66    &0.42 &  0.43 &  -1.88 & 0.82 &G1&Seq    \\               
NGC 7492  &   4.51   &  12.08 &  4.29 &  4.4  & 0.72    &0.12 &  0.37 &  -1.78 & 0.77 &G1&GSE    \\

\hline
\end{tabular}}
\tablefoot{Columns list: cluster name, King and dynamical tidal radii, present-day cluster mass (logarithmic scale), central escape velocity, central concentration, ratio of half-mass radius to dynamical tidal radius, mass-loss fraction, metallicity, relaxation-age ratio, classification type, and possible progenitor system. Among them, $r_{\rm t}^{\rm King}$, $c$ and [Fe/H] are taken from the Harris catalog 
(\citeauthor{1996AJ....112.1487H} \citeyear{1996AJ....112.1487H}, 2010 edition); 
the classification type and progenitor system are derived in this work; 
all other parameters are taken from, or computed using, the globular cluster database used in this study.}
\end{table*}

\begin{table*}
\caption{\raggedright Basic and orbital dynamical parameters of our globular cluster sample.}
\label{tab2}
\centering
\resizebox{\linewidth}{!}{
\begin{tabular}{lccccccccccc}
\hline\hline
Name & $R_\odot$ & $R_{\rm GC}$ & $R_{\rm Peri}$ & $R_{\rm Apo}$ &$a$& $i$ & $e$&$\theta_{\rm R}$&$L_{z}$&$E_{\rm tot}$\\
     & (kpc) & (kpc) & (kpc) &(kpc) &(kpc) &(deg)&&(deg)&($\mathrm{kpc\ }\kms$)
&($10^{11}\ ({\mathrm m\ \mathrm s^{-1})^2}$)
 \\
\hline
NGC 288   &  8.99   &  12.21  &  1.93   &  12.34 &  7.14  &  122.54 & 0.73    &204.31 &  -473.14  &  -1.35   \\                       
NGC 362   &  8.83   &  9.62   &  0.73   &  11.84 &  6.29  &  90.56  & 0.88    &70.77 &  -56.49   &  -1.39   \\                       
Whiting 1 &  30.59  &  35.15  &  35.15  &  61.94 &  48.55 &  74.38  & 0.28    &333.41    &  1729.71  &  -0.54    \\                       
NGC 1261  &  16.40  &  18.28  &  1.65   &  21.08 &  11.37 &  125.12 & 0.85    &255.94 &  -386.99  &  -1.10   \\                       
NGC 1851  &  11.95  &  16.69  &  0.90    &  19.90  &  10.40   &  102.96 & 0.91    &83.07 &  -233.15  &  -1.12   \\                       
NGC 1904  &  13.08  &  19.09  &  0.30    &  19.57 &  9.94  &  77.02  & 0.97    &141.60 &  -39.017  &  -1.13   \\                       
NGC 4147  &  18.54  &  20.74  &  1.83   &  25.27 &  13.55  &  86.33  & 0.86    &82.91 &  -41.65   &  -1.01   \\                       
NGC 4590  &  10.40  &  10.35  &  8.88   &  29.37 &  19.13 &  41.46  & 0.54    &348.81 &  2410.24  &  -0.93   \\                       
NGC 5024  &  18.50  &  19.00  &  9.28   &  22.64 &  15.96  &  75.37  & 0.42    &268.44 &  708.32   &  -1.00   \\                       
NGC 5053  &  17.54  &  18.01  &  10.79  &  17.84 &  14.32 &  76.29  & 0.25    &169.82 &  674.56   &  -1.05   \\                       
NGC 5272  &  10.18  &  12.09  &  5.28   &  15.80  &  10.54  &  57.41  & 0.50    &285.30 &  904.72   &  -1.20   \\                       
NGC 5466  &  16.12  &  16.48  &  6.16   &  47.79 &  26.98 &  107.74 & 0.77    &13.60 &  -869.94  &  -0.72   \\                       
NGC 5634  &  25.96  &  21.84  &  3.08   &  22.18 &  12.63  &  68.01  & 0.76    &210.05 &  479.96   &  -1.05   \\                       
NGC 5897  &  12.55  &  7.41   &  2.45   &  8.83  &  5.64   &  59.82  & 0.56    &76.32 &  392.75   &  -1.52   \\                       
NGC 5904  &  7.48   &  6.27   &  3.25   &  23.44 &  13.35 &  72.77  & 0.76    &350.07 &  364.32   &  -1.09  \\                       
NGC 6205  &  7.42   &  8.64   &  1.51   &  8.73  &  5.12   &  109.53 & 0.71    &207.71 &  -270.26  &  -1.53    \\                       
NGC 6229  &  30.11  &  29.45  &  1.43   &  30.36 &  15.90 &  64.86  & 0.91    &151.96 &  207.40   &  -0.92 \\                       
NGC 6341  &  8.50   &  9.85   &  1.25   &  10.62 &  5.94  &  83.40  & 0.79    &131.00 &  31.49    &  -1.44   \\                       
Terzan 7  &  24.28  &  16.85  &  14.57  &  50.41 &  32.49  &  81.23  & 0.55    &6.73 &  895.89   &  -0.66   \\                       
Arp 2     &  28.73  &  21.39  &  17.77  &  51.38 &  34.58 &  77.85  & 0.49    &10.11 &  1391.55  &  -0.62   \\                       
Terzan 8  &  27.54  &  20.43  &  17.66  &  95.16 &  56.41  &  80.99  & 0.69    &6.02 &  1099.18  &  -0.57   \\                       
NGC 6864  &  20.52  &  14.30  &  1.12   &  16.15 &  8.64  &  56.59  & 0.87    &265.10 &  273.09   &  -1.23   \\                       
NGC 6981  &  16.66  &  12.53  &  0.73   &  22.08 &  11.41 &  108.73 & 0.94    &325.29 &  -13.86   &  -1.09   \\                       
NGC 7089  &  11.69  &  10.54  &  0.88   &  18.74 &  9.81   &  119.89 & 0.91    &45.86 &  -191.31  &  -1.19   \\                       
NGC 7099  &  8.46   &  7.36   &  1.51   &  8.56  &  5.04  &  121.94 & 0.70    &270.18 &  -283.76  &  -1.56  \\                       
Pal 12    &  18.49  &  15.28  &  15.30   &  51.25 &  33.28 &  67.34  & 0.54    &358.42    &  2045.21  &  -0.67   \\                       
Pal 13    &  23.48  &  24.57  &  6.91   &  58.24 &  32.58 &  112.39 & 0.79    &24.61 &  -1233.24 &  -0.64   \\                       
NGC 7492  &  24.39  &  23.57  &  2.81   &  26.03 &  14.42  &  96.40  & 0.80    &257.39 &  -138.74  &  -0.98   \\ 

\hline
\end{tabular}}
\tablefoot{Columns list: cluster name, present heliocentric distance and Galactocentric distance, orbital pericenter and apocenter distances, semi-major axis, orbital inclination, eccentricity, radial angle in action-angle coordinates, angular momentum along the $z$-axis, and total orbital energy. All parameters are taken directly from, or calculated based on, the values provided in the globular cluster database used in this study.}
\end{table*}

\section{Matched filter method}
To detect potential extratidal structures around each globular cluster, we employed the matched filtering technique developed by \citet{2002AJ....124..349R}, which enhances the contrast between cluster members and field stars in CMD space and is effective for detecting low-surface-brightness structures. This technique has been successfully applied in several subsequent studies of tidal features around star clusters (e.g., \citealt{2010AJ....139..606C,2010A&A...522A..71J,2017ApJ...841L..23N,2018MNRAS.474..683C,2018ApJ...862..114S,2020MNRAS.495.2222S,2022ApJ...930...23N,2023ApJ...953..130Y,2024AJ....168..237Z}), demonstrating its effectiveness in tracing surrounding stellar substructures.

According to \citet{2002AJ....124..349R}, at any position on the sky (within a solid angle $d\Omega$), the expected number of stars in any sky bin $(i,j)$ in color-magnitude can be expressed as: 
\begin{equation}
n_{\mathrm{stars},(i,j)} = \alpha  f_{\mathrm{cl},(i,j)} + n_{\mathrm{bg},(i,j)}. 
\label{eq:mf_model}
\end{equation}
Here, $f_{\mathrm{cl}}$ represents the normalized color-magnitude distribution of cluster members and serves as the cluster template in the matched-filter method, while $n_{\mathrm{bg}}$ denotes the background number density distribution in the same bin, which is used as the corresponding background template. The parameter $\alpha$ serves as a scaling factor that quantifies the number (or surface density) of stars following the cluster's CMD distribution within the given sky region. Subsequently, the calculation of this parameter can be derived using the minimum-variance estimator (see Equation 2 in \citealt{2002AJ....124..349R}).

\begin{figure*}[htbp]
\center
\includegraphics[width=0.19\textwidth]{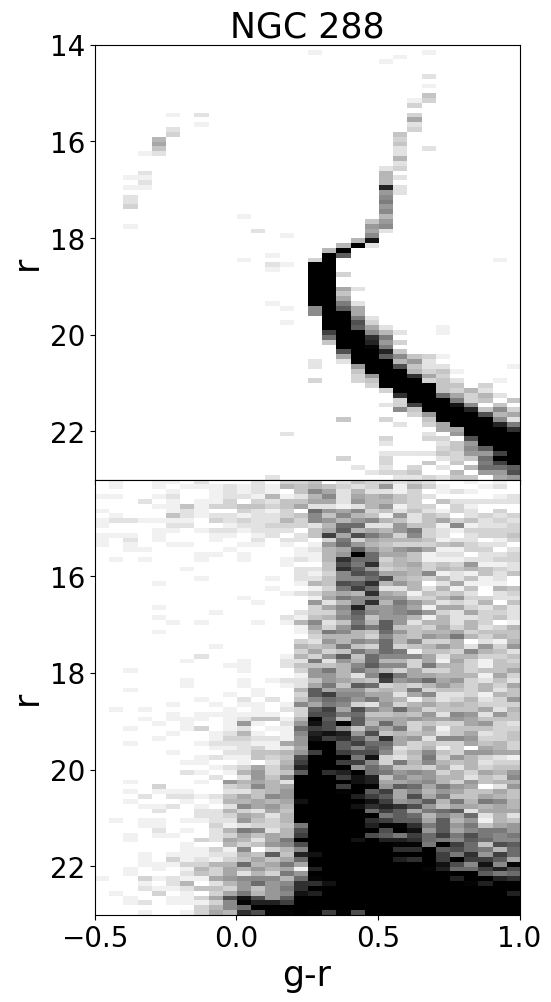}
\includegraphics[width=0.19\textwidth]{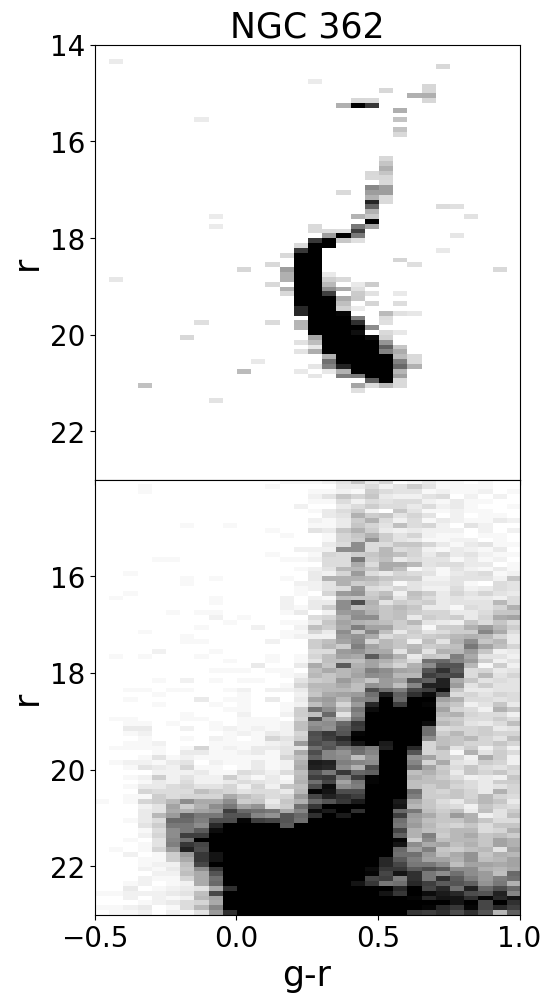}
\includegraphics[width=0.19\textwidth]{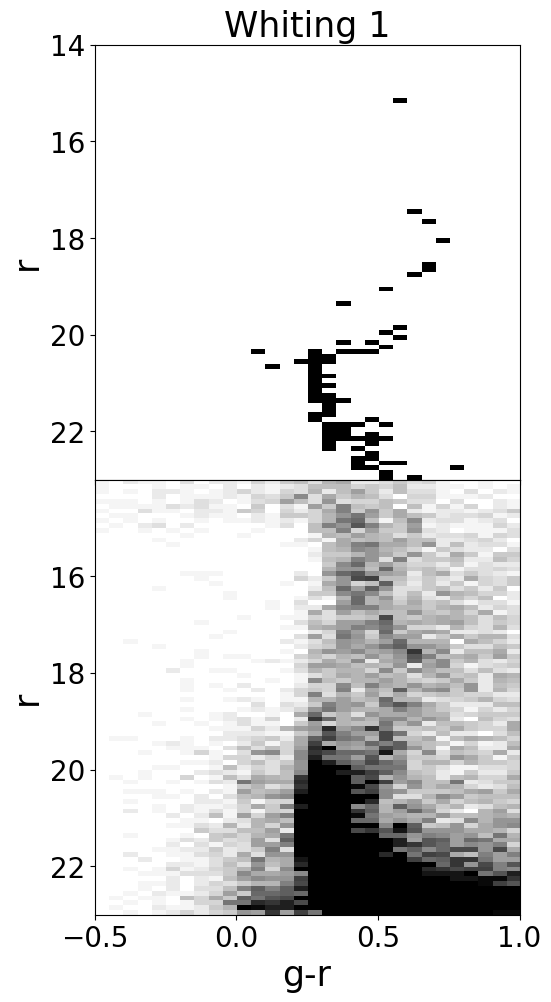}
\includegraphics[width=0.19\textwidth]{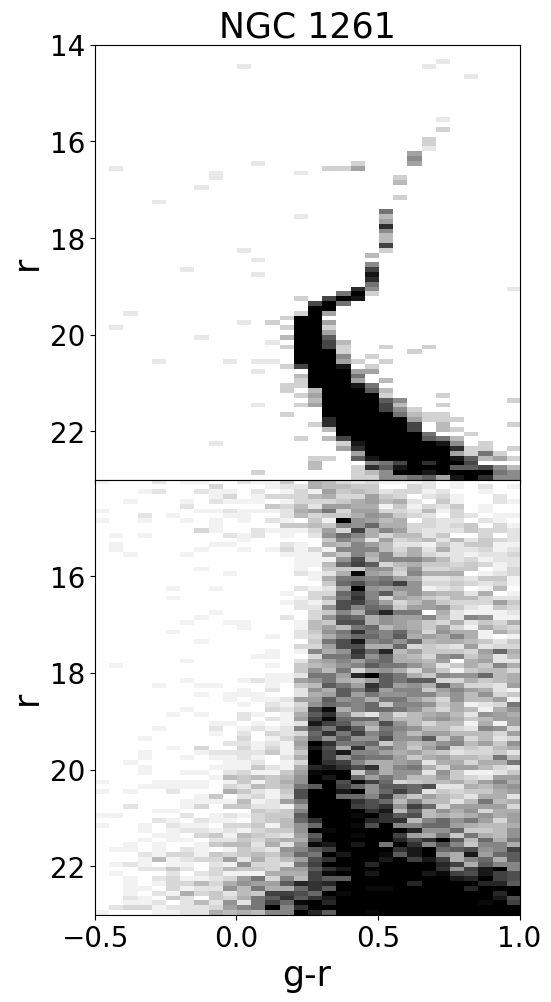}
\includegraphics[width=0.19\textwidth]{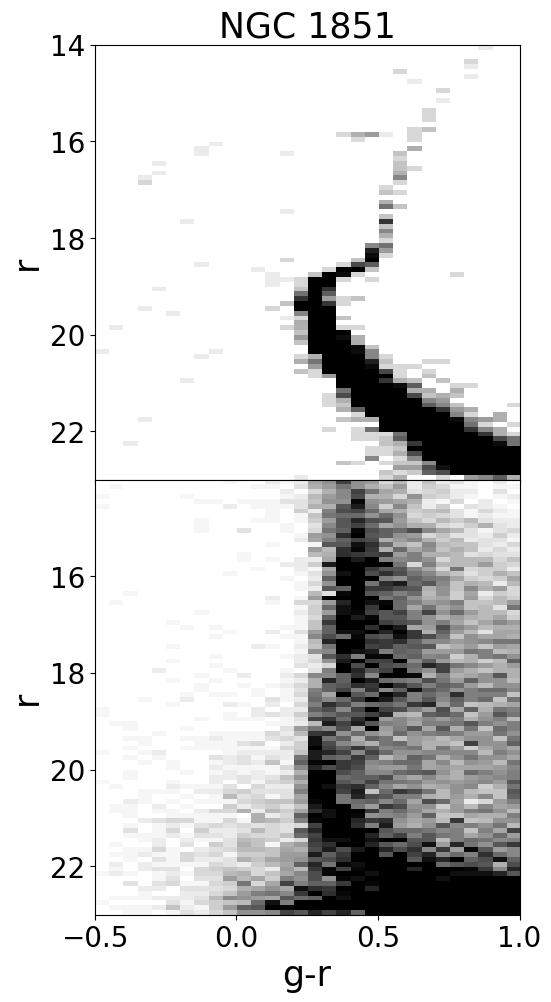}

\center
\caption{Color-magnitude diagrams for part of the cluster sample.
Each panel displays the template Hess diagram (top) and the global background Hess diagram (bottom). CMDs for the remaining clusters
are presented in Fig. \ref{CMDappendix}.}
\label{CMD}
\end{figure*}

Based on this model, we first constructed the cluster CMD template ($f_{\mathrm{cl}}$) used in the matched-filtering analysis. To do so, we generated a Hess diagram for the cluster region using bin sizes of $\Delta(g - r)=0.05$ and $\Delta r=0.1$ mag. In most cases, the CMD was built in the $(g - r, r)$ plane. However, for a few clusters lacking $r$-band photometry, such as Ter 7, Ter 8, and Pal 12, the $(g - i, i)$ combination was adopted instead. In order to reduce the effects of crowding and incompleteness in the cluster core, stars within the inner 1{\arcmin}-3{\arcmin} were excluded from the template construction. For clusters with relatively small tidal radius and more diffuse stellar distributions (e.g., Whiting 1 and Pal 13), no such exclusion was applied. Consequently, the cluster template was constructed from stars located between this exclusion zone (when applied) and the cluster's tidal radius (from \citeauthor{1996AJ....112.1487H} (\citeyear{1996AJ....112.1487H}), 2010 edition). In addition, to reduce field-star contamination and obtain a clean cluster CMD template $f_{\mathrm{cl}}$, we selected stars with $-0.5 < g - r < 1.0$ and $14 < r < 23$ mag, and subtracted a local background derived from a comparing field in an annular region centered on the cluster, typically located between about $1^{\circ}$ and $1.5^{\circ}$ from the cluster center. The stellar density in this annular region was normalized to the area of the cluster region and subtracted from the cluster Hess diagram, yielding a background-subtracted CMD template (see the upper panels of each Hess diagram in Fig.~\ref{CMD}). In contrast, the global background CMD template ($n_{\mathrm{bg}}$) required for the matched-filtering analysis was constructed from four rectangular control regions ($1^{\circ} \times 1^{\circ}$) located at different position angles far away from the cluster, each more than $2^\circ$ from the cluster center. The Hess diagrams of these regions were then averaged to produce a uniform global background model (see the lower panels of each Hess diagram in Fig.~\ref{CMD}). 

Finally, the matched-filter procedure yields the quantity $\alpha$, which represents the number of stars that pass the matched filter within a given spatial bin. Using these $\alpha$ values, we then constructed a spatial density map to reveal possible tidal structures. However, if the spatial bins are too large, thin tidal features may be smeared out, whereas overly small bins introduce noise and may generate spurious features. Therefore, after testing a range of spatial binning scales and visually inspecting the results, we ultimately adopted a uniform grid with an angular resolution of $0.02^{\circ}\times0.02^{\circ}$. To evaluate the statistical significance of the detected structures, we computed a signal-to-noise (S/N) map. For each spatial bin, we subtracted the mean background level (estimated from the outer regions excluding the cluster) from its $\alpha$ value and then divided the result by the standard deviation of the background. This final S/N map reveals potential extratidal features with enhanced contrast (see Fig. \ref{figure2}).

\section{Result}
\subsection{Detection of tidal structures}

\begin{figure*}[htbp]
\center
\includegraphics[width=0.32\textwidth]{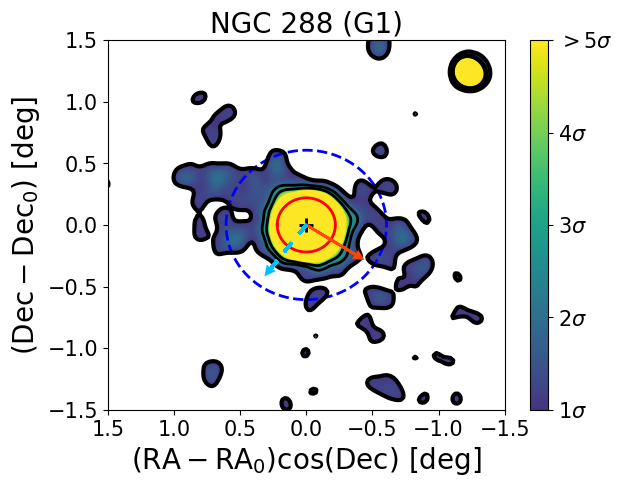}
\includegraphics[width=0.32\textwidth]{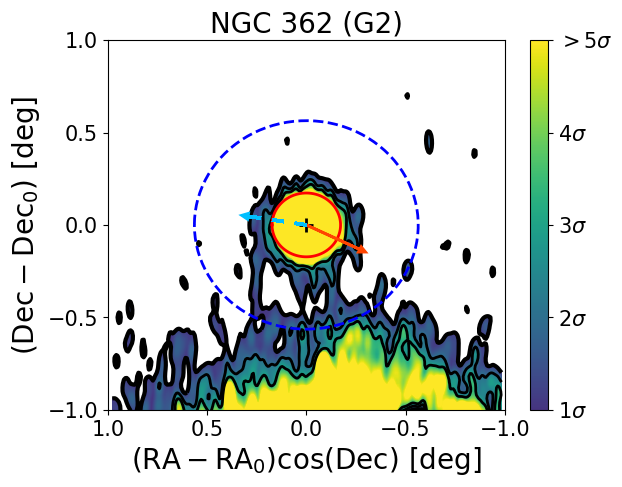}
\includegraphics[width=0.32\textwidth]{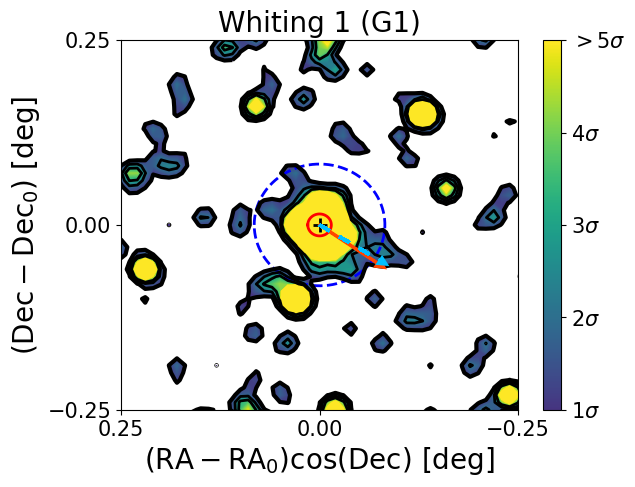}

\center
\caption{Signal-to-noise ratio distributions for part of the cluster sample, derived from the matched-filter output. The red circle indicates the King tidal radius (${r_{\rm t}}^{\rm King}$) from the Harris catalog, while the dashed blue circle represents the dynamical tidal radius (${r_{\rm t}}^{\rm dyn}$) from the globular cluster database. The solid orange arrow points toward the Galactic center, and the dashed light-blue arrow indicates the direction of the cluster's orbital motion over a short timescale. The black contours correspond to significance levels from 1$\sigma$ to 3$\sigma$, with the outermost 1$\sigma$ contour shown in bold. S/N distributions for
the remaining clusters are shown in Fig. \ref{figure2appendix}.}
\label{figure2}
\end{figure*}

Based on the matched filtering technique, we constructed S/N maps for all clusters in our sample (see Fig. \ref{figure2}). The maps are presented in a projected coordinate system centered on each cluster, with axes defined as $\Delta \mathrm{RA}\times \cos(\mathrm{Dec})$ and $\Delta \mathrm{Dec}$, ensuring consistent spatial analysis across different celestial positions. The cluster centers were adopted from the Galactic globular cluster database \footnote{people.smp.uq.edu.au/HolgerBaumgardt/globular/}, which is extensively used in this study. This database, developed by Baumgardt and collaborators, provides a comprehensive set of physical and dynamical parameters for Galactic GCs. These parameters were derived by combining a compilation of observational data, including ground-based radial velocities, Gaia DR3 proper motions, and HST-based stellar mass functions, with literature data, and by performing N-body modeling and orbital integrations \citep[see, e.g.,][etc.]{2017MNRAS.464.2174B,  2018MNRAS.478.1520B, 2021MNRAS.505.5978V, 2021MNRAS.505.5957B, 2023MNRAS.521.3991B}. Further details can be found in the database.

We adopted a uniform spatial resolution of $0.02^\circ$ ($=1\arcmin.2$) for all clusters except for Whiting 1. This cluster is a relatively small system, with a King-model tidal radius of only $0\arcmin.89$ according to the Harris catalog (\citeauthor{1996AJ....112.1487H} \citeyear{1996AJ....112.1487H}, 2010 edition), and it lies at a heliocentric distance of $30.59$ kpc \citep{2021MNRAS.505.5957B}. As a result, its projected angular extent on the sky is limited. To better resolve its structural features, we adopted a finer spatial resolution of $0.01^\circ$ ($= 0\arcmin.6$) for this target. In addition, we applied Gaussian smoothing to the matched-filter S/N maps to enhance the visibility of extended structures while suppressing small-scale noise. Because the clusters in our sample differ substantially in distance, apparent angular size, and the spatial distribution of their member stars, a single smoothing scale is not appropriate. We therefore adopted $\sigma$ = 1 pixel as the default smoothing kernel. The kernel size was then adjusted through systematic visual inspection of the resulting S/N maps to identify the smallest kernel that effectively reduces noise without oversmoothing genuine features. Such inspections help effectively reduce the risk of smoothing out real features or missing potential tidal structures from insufficient smoothing. In practice, the smoothing kernels adopted for our sample range from $\sigma$=1 to 5 pixels, depending on the cluster's properties.

To investigate the properties of extratidal features among our sample clusters, we analyzed the matched-filtered S/N maps and adopted a classification scheme similar to that of \citet{2020A&A...637L...2P}. Specifically, we overlaid two reference tidal radii on each map: the King tidal radius (${r_{\rm t}}^{\rm King}$) taken from the Harris catalog (\citeauthor{1996AJ....112.1487H} \citeyear{1996AJ....112.1487H}, 2010 edition), which is derived from surface density fitting under the King model \citep{1962AJ.....67..471K}, and the dynamical tidal radius (${r_{\rm t}}^{\rm dyn}$) from the globular cluster database \footnote{people.smp.uq.edu.au/HolgerBaumgardt/globular/parameter.html}, which represents the present-day limiting radius $r_{\rm L}(F)$ calculated following the expression of \citet{2013ApJ...764..124W}: 
\begin{equation}
r_{\rm L}(F) = r_{\rm t}(R_{\rm Peri})\left(1 + aF \exp(b \cdot e)\right).
\end{equation}
Here, $r_{\rm t}(R_{\rm Peri})$ is the tidal radius at pericenter, $e$ is the orbital eccentricity, and $F$ is the orbital phase defined as
\begin{equation}
F = \frac{R_{\rm GC} - R_{\rm Peri}}{R_{\rm Apo} - R_{\rm Peri}},
\end{equation}
where $R_{\rm GC}$ is the current Galactocentric distance of the cluster, and $R_{\rm Apo}$ and $R_{\rm Peri}$ denote the apocenter (maximum) and pericenter (minimum) distances along its orbit, respectively. The coefficients are $a = 0.17 \pm 0.03$ and $b = 4.1 \pm 0.2$ \citep{2013ApJ...764..124W}. Compared to the King tidal radius, the dynamical tidal radius provides a physically motivated boundary that accounts for orbital dynamics and gravitational interactions with the Galactic potential. 

All clusters in our sample were classified into three groups according to their structural properties with respect to $r_{\rm t}^{\rm King}$ and ${r_{\rm t}}^{\rm dyn}$: G1 clusters exhibit coherent, tail-like extensions aligned along a preferred direction, with detectable overdensities (typically reaching $\mathrm{S/N}\ge1\sigma$) that extend beyond the dynamical tidal radius ($r_{\rm s}>r_{\rm t}^{\rm dyn}$); G2 clusters exhibit overdensities with $\mathrm{S/N}\ge3\sigma$ extending beyond the King tidal radius ($r_{\rm s}>r_{\rm t}^{\rm King}$), while being overall confined within the dynamical tidal radius ($r_{\rm s}< r_{\rm t}^{\rm dyn}$). These extensions typically appear as diffuse envelopes, irregular overdensities, or clumpy halo-like features, and may in some cases extend beyond $r_{\rm t}^{\rm dyn}$. In addition, clusters that display tail-like structures beyond $r_{\rm t}^{\rm King}$ but still remain fully contained within $r_{\rm t}^{\rm dyn}$ are also included in this group; G3 clusters show no overdensities with $\mathrm{S/N}\ge3\sigma$ beyond the King tidal radius ($r_{\rm s}< r_{\rm t}^{\rm King}$), and whose outer stellar density profiles are consistent with the truncation expected from a King model.

It is worth noting that for G1 clusters, we allowed overdensities down to the 1$\sigma$ level because the stellar surface density beyond $r_{\rm t}^{\rm dyn}$ is intrinsically low, meaning that genuine tidal extensions may in some cases be detected only with modest significance. Despite their lower S/N, these features still exhibit coherent, directionally aligned, tail-like structures, which remain reliable indicators of tidal disturbance. The distinction between G1 and G2 is primarily based on the stellar density map. In cases where the classification is unclear, the angular density distribution is used as a supplementary tool. The angular density distribution is obtained by dividing the region around the cluster into 36 sectors (pie-shaped regions) and calculating the total number density (here is total $\alpha$) within each sector. Tail-like structures exhibit density peaks in one or more sectors, while diffuse or irregular envelopes show a more uniform density across the sectors. In addition, to ensure the reliability of our classification, we compared our results with those from previous studies and discussed the differences between them. Based on our results, we classified 12 clusters as G1, nine clusters as G2, and seven clusters as G3. We also report the first detection of an extended tidal structure around the Sagittarius-associated cluster Terzan 7 (\citeauthor{2019A&A...630L...4M} \citeyear{2019A&A...630L...4M}, \citeyear{2025RNAAS...9...64M} update; \citealt{2020MNRAS.493..847F}). A detailed analysis of each cluster is presented in Appendix \ref{sec:appendix_clusters}.

\subsection{Dependence of tidal features on intrinsic properties}

To investigate whether any internal parameters can effectively predict the presence or morphology of tidal structures in GCs, we selected a representative set of physical quantities that are commonly associated with the structural and dynamical evolution of clusters. These include the present-day cluster mass ($\log M$), central escape velocity ($v_{\rm esc}$), central concentration ($c$), ratio of half-mass radius to dynamical tidal radius ($r_{\rm h,m}/r_{\rm t}^{\rm dyn}$), mass-loss fraction ($M_{\rm dis}/M_{\rm init}$), metallicity ($\rm [Fe/H]$), and relaxation-age ratio ($\log({\rm age}/{T}_{\rm rh})$). Except for $c$ and $\rm [Fe/H]$, which are taken from the Harris catalog (\citeauthor{1996AJ....112.1487H} \citeyear{1996AJ....112.1487H}, 2010 edition), all other quantities are derived from the base data provided by the globular cluster database \footnote{people.smp.uq.edu.au/HolgerBaumgardt/globular/} \footnote{people.smp.uq.edu.au/HolgerBaumgardt/globular/parameter.html}. The mass-loss fraction is calculated following the prescription of \citet{2020A&A...637L...2P}.

Fig. \ref{para1} shows the corner plot of the selected intrinsic parameters, with clusters color-coded by morphology (G1 in red, G2 in blue, and G3 in black). The diagonal panels display distributions of individual parameters, while the lower-triangular panels present the pairwise correlations. To further analyze whether the distributions of each parameter differ among the groups, we performed two-sample Kolmogorov-Smirnov (K-S) tests on the distribution of each parameter across different groups. It should be noted that, due to the limited sample size, the K-S tests serve only as supporting evidence rather than definitive conclusions. Table \ref{tab3} summarizes the test results, where we report only the false-discovery-rate (FDR) corrected $p$-values ($q$). We adopt $q < 0.05$ as the threshold for statistical significance, indicating that the two samples differ significantly. Pal 12 is highlighted as an outlier because its King tidal radius exceeds the dynamical radius and it consistently appears displaced across multiple parameter spaces; it is therefore marked with an open symbol in Fig. \ref{para1} and excluded from further analyses in this subsection.

\subsubsection{Possible group-sensitive parameters}
From analysis of Fig. \ref{para1}, we observed a positive correlation between $\log_{\rm 10}(M/M_\odot)$ and $v_{\mathrm{esc}}$. Clusters in Group G1, which have lower mass, also exhibit lower escape velocity, whereas clusters in Group G2 and G3 generally occupy higher values in both parameters. This behavior is consistent with theoretical expectations, as the total mass determines the depth of a cluster's gravitational potential well and shallower gravitational potential wells make clusters more susceptible to external perturbations and easy to escape \citep{2003MNRAS.340..227B}. The two-sample K-S tests reveal significant differences between G1 and G2 ($q=0.02$ for $\log M$ and $q=0.04$ for $v_{\mathrm{esc}}$) as well as between G1 and G3 ($q=0.01$ for $\log M$ and $v_{\mathrm{esc}}$). In contrast, no significant differences are found between G2 and G3 ($q>0.1$). These results show that G1 clusters are clearly distinct from the other two groups. The absence of significant differences between G2 and G3 suggests that G2 clusters may occupy an intermediate state rather than forming a fully separate class. In this sense, G2 may represent a transitional population.

The parameter $c = \log(r_{\rm t}/r_{\rm c})$ reflects the central compactness of the cluster. Examining the distribution of $c$ across the groups, we find a pattern similar to that of $\log_{\rm 10}(M/M_\odot)$ and $v_{\mathrm{esc}}$: clusters in G1 predominantly occupy lower values, whereas those in G2 and G3 are concentrated at higher values. This trend is consistent with theoretical expectations. \cite{1990ApJ...351..121C} found that clusters with lower central concentration and flatter mass functions are more easily disrupted due to cluster expansion driven by stellar evolution and tidal mass loss. Therefore, more compact clusters are likely less susceptible to developing tidal structures. Moreover, Fig. \ref{para1} reveals clear positive correlations between $\log_{\rm 10}(M/M_\odot)$ and $c$, as well as between $v_{\mathrm{esc}}$ and $c$, indicating that more massive clusters generally have higher central concentrations and stronger gravitational binding, making them less prone to forming tidal tails. Two-sample K-S tests further indicate significant differences between G1 and G2/G3 ($q=0.01$), while G2 and G3 show similar distributions ($q=0.27$), supporting the interpretation of G2 as a transitional population.

Defined as the ratio between the half-mass radius and the dynamical tidal radius, $r_{\rm h,m}/r_{\rm t}^{\rm dyn}$ measures how tidally filled a cluster is. From Fig. \ref{para1}, the distribution of this parameter in G1 differs from that in G2/G3: G1 spans the entire parameter space, whereas G2/G3 are confined to lower values. The K-S tests further confirm a significant difference between G1 and G3 ($q=0.04$), while no significant difference is found between G1 and G2 ($q=0.09$) or between G2 and G3 ($q=0.18$). In addition, the correlation plots show that this parameter is noticeably anticorrelated with $\log_{\rm 10}(M/M_\odot)$, $v_{\mathrm{esc}}$, and $c$.  This can be explained theoretically, since more massive or more compact clusters are more resistant to tidal filling and subsequent disruption \citep{1990ApJ...351..121C}.

\subsubsection{Group-insensitive parameters}
\begin{figure*}
\center

\includegraphics[width=0.7\textwidth]{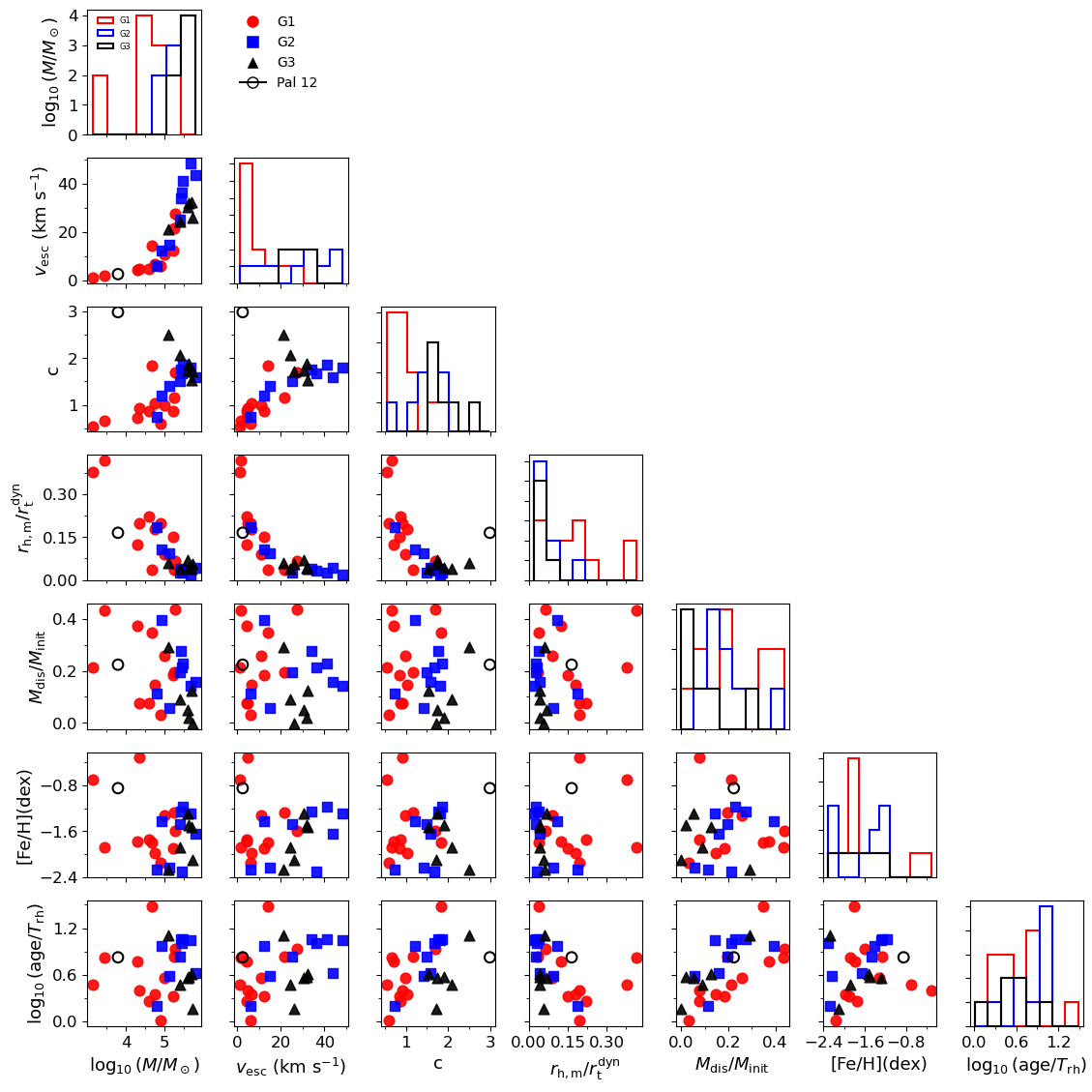}

\center
\caption{Corner plot of intrinsic parameters. The selected parameters include total mass ($\log M$), escape velocity ($v_{\rm esc}$), central concentration ($c$), ratio of half-mass radius to dynamical tidal radius ($r_{\rm h,m}/r_{\rm t}^{\rm dyn}$), mass-loss fraction ($M_{\rm dis}/M_{\rm init}$), metallicity ($\rm {[Fe/H]}$) and relaxation-age ratio ($\log(\mathrm{age}/T_{\rm rh})$). The diagonal panels show the one-dimensional histograms of each parameter for the three morphological groups, while the off-diagonal panels display the corresponding two-dimensional scatter plots for all parameter pairs. Clusters are color-coded by morphological classification: red for G1, blue for G2, and black for G3. Pal 12 is highlighted with open marker.}
\label{para1}
\end{figure*}

The ratio $M_{\rm dis}/M_{\rm init}$ quantifies the fractional mass loss from the cluster's initial mass to its current mass. From Fig. \ref{para1}, we also observe a clear distributional difference between G1 and G3: Clusters in G1 tend to lose a larger fraction of their member stars, making them more prone to developing tidal structures, whereas those in G3 lose fewer stars and are therefore less likely to form such features. G2 appears to represent a transitional population between the two. This parameter also exhibits weak correlations with several physical quantities: clusters with higher $\log_{\rm 10}(M/M_\odot)$, larger $v_{\mathrm{esc}}$, or higher $c$ tend to show lower $M_{\rm dis}/M_{\rm init}$, whereas more tidally filled systems generally display higher values of $M_{\rm dis}/M_{\rm init}$. In addition, its positive correlation with $\log(\mathrm{age}/T_{\rm rh})$, suggests that clusters that have lost more stars also experience more dynamical relaxation. However, the K-S tests do not reveal any significant differences among the three groups ($q>0.1$). We speculate that this may be influenced by the G2 clusters, as these systems also exhibit rich extratidal features and may have experienced a comparable degree of mass loss. In addition, the coupling between this parameter and the clusters' orbital dynamical properties may introduce further scatter, thereby weakening the statistical separation. Therefore, although this parameter carries meaningful physical correlations, we temporarily classify it as group-insensitive.

The quantity $\rm [Fe/H]$ is widely used as a tracer of the formation environment and epoch, and may also indicate whether a cluster originated in an external satellite galaxy. Metal-poor clusters are often linked to accreted populations and may have undergone more complex interaction histories \citep{2019A&A...630L...4M, 2020MNRAS.493..847F}, potentially shaping their outer structures. In our analysis, however, we do not find significant differences in its distribution across different groups. \citet{2025AJ....170..157K} recently reported a weak correlation between metallicity and tidal morphology, but their sample is restricted to relatively nearby and massive clusters, which could bias the inferred relation.

The logarithmic ratio $\log_{\rm 10}(\mathrm{age}/T_{\rm rh})$ between the cluster's age and its half-mass relaxation time indicates the degree of dynamical relaxation. From the parameter distributions in Fig. \ref{para1}, there is no obvious difference among the groups. The pairwise correlation plots reveal the previously mentioned correlation between $\log(\mathrm{age}/T_{\rm rh})$ and $M_{\rm dis}/M_{\rm init}$.
In the study by \citet{2019MNRAS.489.4367P}, it was proposed that GCs with relatively short relaxation times often have already undergone substantial mass loss. Such depletion, typically caused by tidal stripping or evaporation, weakens the system's gravitational binding and accelerates its internal dynamical evolution, thereby increasing the likelihood of further stellar escape. This interpretation is consistent with both theoretical and numerical studies: once mass loss begins, the relaxation time shortens, which enhances the efficiency of two-body interactions and accelerates core evolution. For example, \citet{2010MNRAS.409..305L} pointed out that external tidal perturbations and the gradual depletion of loosely bound stars can drive a positive feedback loop between mass loss and structural evolution, while \citet{2008MNRAS.389L..28G} showed that clusters evolving within the realistic Galactic tidal field exhibit a weaker scaling between dissolution time and relaxation time than in isolation ($t_{\rm dis} \propto t_{\rm rh}^{3/4}$), highlighting the accelerating role of tidal stripping. Thus, a short relaxation time may not only be interpreted as the dynamical consequence of the preferential loss of outer stars, but also as a potential precursor to the emergence of extended tidal features. In our analysis, the parameters $M_{\rm dis}/M_{\rm init}$ and $\log(\mathrm{age}/T_{\rm rh})$ show a certain degree of linear correlation, which may reflect the influence of such evolutionary processes. However, no significant clustering or separation is observed either in this parameter space or in those involving $\log(\mathrm{age}/T_{\rm rh})$. This suggests that while relaxation time and its related ratios play an important role in the dynamical evolution of GCs, they alone are insufficient to reliably characterize the presence or extent of tidal structures, particularly given the diverse environments and evolutionary stages represented in our sample.

These results indicate that internal parameters, particularly those linked to the depth of the gravitational potential well, play an important role in shaping the formation and characteristics of tidal structures, even though some overlap between groups remains. At the same time, tidal interactions are inherently time-dependent, and the strength and frequency of external perturbations can vary substantially along a cluster's orbit. Therefore, the development and morphology of tidal structures are shaped not only by internal dynamics but also by the broader orbital context and external environment. To further explore these effects, we now turn to an analysis of orbital dynamical parameters.

\begin{table}
\caption{\raggedright Two-sample K-S test results for internal and orbital parameters among different cluster groups.}
\label{tab3}
\centering
\begin{tabular}{lcccccc}
\hline\hline
Parameter && $q$&\\
      &(G1 vs. G2) & (G1 vs. G3)&(G2 vs. G3)\\
\hline
$\log M$   & 0.02 & 0.01&0.41 \\
$v_{\rm esc}$&  0.04 & 0.01&0.18 \\
$c$ &  0.01 & 0.01&0.27 \\
$r_{\rm h,m}/r_{\rm t}^{\rm dyn}$  &  0.09&0.04 & 0.18 \\
$M_{\rm dis}/M_{\rm init}$ &  0.93 & 0.17&0.17 \\
$\rm [Fe/H]$ &  0.84 & 0.96&0.84 \\
$\log(\mathrm{age}/T_{\rm rh})$ &  0.24 & 0.96&0.24 \\
$R_{\rm Peri}$ &  0.02 & 0.76&0.02 \\
$e$ &  0.06 & 0.25&0.04 \\
$\theta_{\rm R}$ &  0.27 &0.04&0.09 \\ 
$R_{\rm Apo}$ &  0.17 & 0.14 &0.78 \\
$a$     & 0.34 &0.38&0.78 \\
$i$     &  0.78 &0.76&0.78 \\

\hline
\end{tabular}
\tablefoot{Col.1: parameter name; Col.2-4: results of the pairwise K-S tests between the three groups (G1 vs. G2, G1 vs. G3, G2 vs. G3), reported as the FDR-adjusted $p$-values ($q$).}
\end{table}

\subsection{Influence of orbital parameters on tidal feature formation}

To further investigate the orbital dynamical mechanisms that may drive the formation of tidal structures in GCs, we analyzed a set of orbital parameters that are commonly linked to tidal interactions with the Galactic potential. Specifically, we considered the pericentric distance ($R_{{\rm Peri}}$), apocentric distance ($R_{{\rm Apo}}$), orbital semi-major axis ($a$), orbital inclination angle ($i$), orbital eccentricity ($e$), and the radial angle in action-angle coordinates ($\theta_{\rm R}$). These quantities collectively describe the cluster's trajectory through the Galaxy and its exposure to varying tidal forces over time. All basic data are taken from the globular cluster database\footnote{people.smp.uq.edu.au/HolgerBaumgardt/globular/orbits.html}. The semi-major axis, eccentricity, and inclination are computed following the formulae in \citet{2019ApJ...882...98P}. The radial angle ($\theta_{\rm R}$) in action-angle coordinates is calculated using the \textit{Galpy} package, adopting the \textit{MWPotential2014} Galactic potential \citep{2015ApJS..216...29B}. This parameter not only characterizes the current position of the cluster along its orbit but also indicates its direction of motion between the pericenter and apocenter. Specifically, when $0^{\circ} < \theta_{\rm R} < 180^{\circ}$, the cluster is moving outward from pericenter toward apocenter, whereas when $180^{\circ} < \theta_{\rm R} < 360^{\circ}$, it is moving inward from apocenter back toward pericenter.

\subsubsection{Possible group-sensitive parameters}
\begin{figure*}
\center

\includegraphics[width=0.7\textwidth]{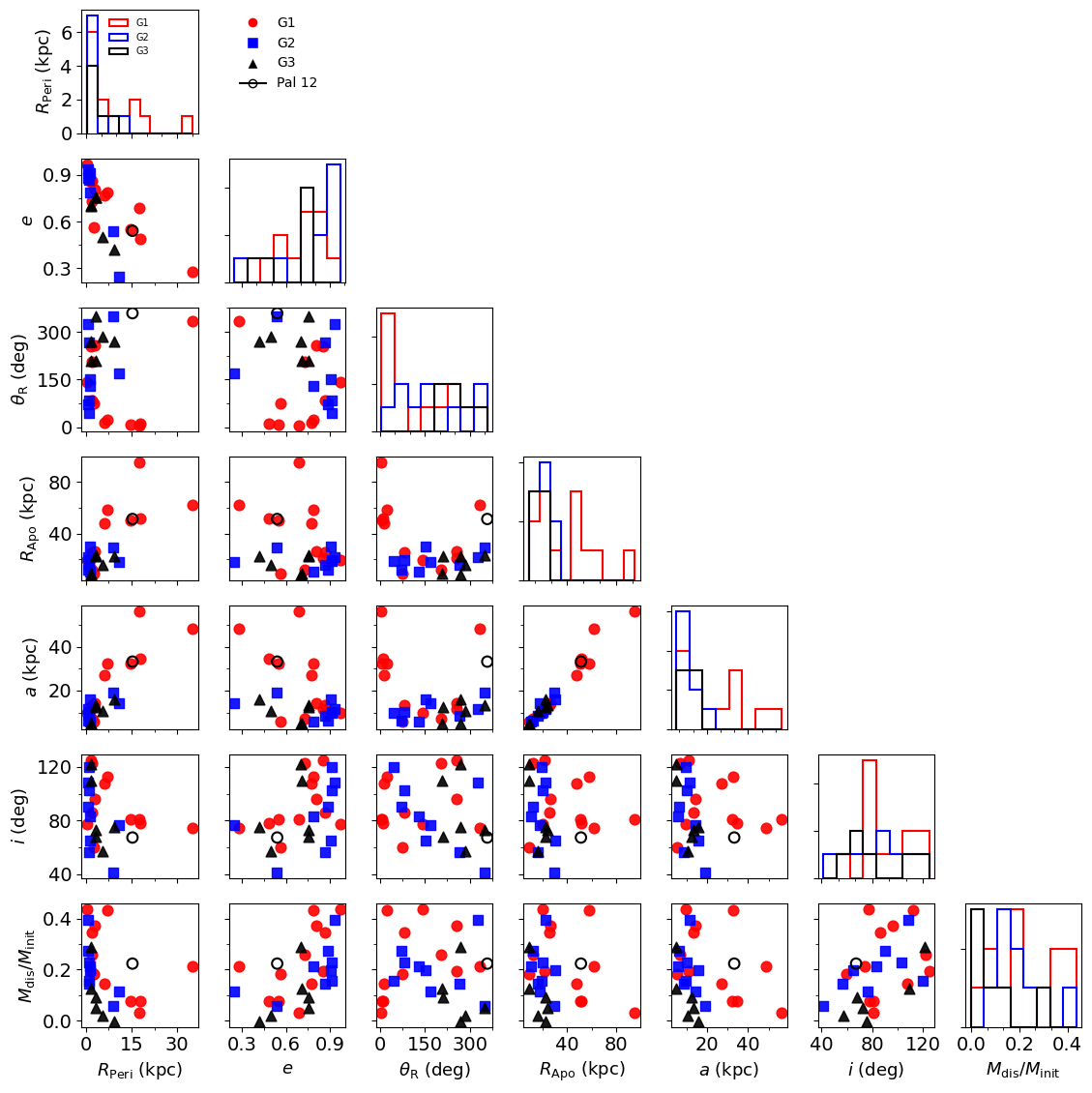}

\center
\caption{Corner plot of orbital parameters. The parameters included are pericenter distance ($R_{\mathrm{Peri}}$), apocenter distance ($R_{\mathrm{Apo}}$), orbital semi-major axis ($a$), inclination angle($i$), orbital eccentricity($e$), and radial angle in action-angle coordinates ($\theta_{\rm R}$). The mass-loss fraction ($M_{\rm dis}/M_{\rm init}$) is also included to test its potential coupling with orbital properties. The color-coding of morphological classifications, the use of symbols, and the presentation of diagonal/off-diagonal panels are the same as in Fig. \ref{para1}.}
\label{para2}
\end{figure*}

Fig. \ref{para2} presents the corner plot of the selected orbital parameters, with clusters colored and symbolized in the same way as in Fig. \ref{para1}. The parameter $M_{\rm dis}/M_{\rm init}$ is also included to test whether orbital properties are coupled with long-term mass depletion. From the distributions shown in Fig. \ref{para2}, we find $R_{\rm Peri}$, $R_{\rm Apo}$, $a$, $i$, $e$ and $\theta_{\rm R}$ exhibit varying degrees of differences between Groups G1 and G3, while Group G2 appears to act as a transitional population. However, the K-S test reveals significant differences only in $R_{\rm Peri}$, $e$ and $\theta_R$ with $q < 0.05$, suggesting that these three parameters may be group-sensitive.

In the two-dimensional orbital parameter space, several notable correlations emerge. Clusters with smaller pericentric distances ($R_{\rm Peri}$) generally exhibit higher mass-loss fractions, consistent with the expectation that stronger tidal forces near the Galactic center accelerate mass depletion. This relation resembles a rough power-law trend, although some G3 clusters with small $R_{\rm Peri}$ show relatively low mass-loss fractions, suggesting that additional mechanisms may mitigate their long-term stripping. Clear geometric correlations are also evident among $R_{\rm Peri}$, $R_{\rm Apo}$, and $a$, but these simply reflect basic orbital mechanics. Clusters on highly eccentric ($e \gtrsim 0.7$) and strongly inclined ($i \gtrsim 70^{\circ}$) orbits are more frequently found among G1, and such orbital configurations are typically associated with enhanced mass loss, even though some overlap exists among different morphological groups. \citet{2014MNRAS.445.1048W} argued that repeated passages of GCs through the Galactic disk can amplify tidal disruption and increase mass loss. Building on this, \citet{2019ApJ...882...98P} proposed that clusters with higher eccentricities and inclinations are more likely to undergo repeated disk crossings, thereby increasing their susceptibility to tidal depletion. \citet{2020A&A...637L...2P} further showed that clusters on highly eccentric ($e \gtrsim 0.8$) and strongly inclined ($\approx \pm 70^{\circ}$) orbits tend to lose more mass than those on nearly circular, low-inclination orbits. Our results are broadly consistent with this framework, while also indicating that not all clusters on such orbits necessarily display prominent tidal features.

Moreover, \citet{2018MNRAS.474.2479B} suggested that GCs exhibiting tidal tails are typically in a near-dissolution phase and are more likely to be found near apocenter. In our sample, a considerable fraction of G1 clusters are located in the outward-moving orbital phase between pericenter and apocenter (with $\theta_{\rm R}$ between 0$^{\circ}$ and 180$^{\circ}$), consistent with the expectation that enhanced tidal stripping occurs during pericentric passages. However, we do not find clear evidence that these clusters are preferentially concentrated near apocenter. Although $\theta_{\rm R}$ and the mass-loss fraction show no strong correlation, G1 clusters nonetheless tend to populate this outward-moving phase, suggesting that the tidal stripping experienced during recent pericentric passages may contribute to the formation of their observed extratidal structures.

Taken together, our results indicate that certain orbital characteristics, such as smaller $R_{\rm Peri}$, higher $e$, and specific $\theta_{\rm R}$, may be associated with enhanced tidal mass loss, and that these three parameters are group-sensitive. These findings highlight that, beyond intrinsic properties, the orbital configuration of GCs also plays a key role in determining both the extent of their mass loss and the detectability of their tidal structures.

\subsubsection{Tidal feature orientations versus orbital direction}
Meanwhile, to investigate whether the orientation of tidal structures is related to the orbital direction of the clusters, we overlaid the orbital directions on the matched-filter S/N maps shown in Fig.~\ref{figure2}. The sky-blue arrows indicate the short-period orbital directions derived from orbital integration, while the orange arrows mark the directions toward the Galactic center. The orbital integration was performed using the \textit{Galpy} package with the  \textit{MWPotential2014} gravitational model \citep{2015ApJS..216...29B}. This directional information provides additional context for interpreting whether the detected features are shaped primarily by orbital motion or Galactic tides.

Our analysis reveals three distinct alignment patterns of tidal features among the clusters in our sample. In some systems (e.g., Whiting 1, NGC 5466), the extended structures broadly align with the orbital path, consistent with theoretical expectations that tidal stripping occurs preferentially along the orbit, producing leading and trailing tails \citep[e.g.,][]{2007ApJ...659.1212M,2010MNRAS.401..105K}. In others (e.g., NGC 1261, NGC 7492), the extensions are partially oriented toward the Galactic center, suggesting that the external tidal field has played a key role, possibly through recent disk-crossing events or similar perturbations \citep[e.g.,][]{1999A&A...352..149C,2011MNRAS.416..393B}. A third group of clusters (e.g., NGC 1851, NGC 6981) displays more symmetric, envelope-like morphologies with no clear alignment. In some cases, certain systems show visible tidal tails that are not aligned with the orbital direction (e.g., NGC 1904, NGC 4147), which may be related to recent pericentric passages \citep{2025A&A...693A..69A}.

Overall, while some clusters do exhibit tidal features aligned with either the orbital direction or the Galactic center, the morphological diversity observed across the sample suggests that the formation and structure of these features are far more complex than initially expected.

\subsection{Environmental independence of tidal morphologies}

To further assess the influence of the Galactic environment on the formation of tidal structures, we investigated several additional parameters. Specifically, we examined the pericentric distance ($R_{\rm Peri}$), the radial orbital period ($P_{\rm R}$), the time elapsed since the last pericentric passage ($T_{\rm Peri}$) and since the most recent disk crossing ($T_{\rm Disk}$), as well as potential associations with major accretion events that may have influenced the clusters' dynamical evolution and external tidal environment.

\begin{figure}
\center

\includegraphics[width=0.47\textwidth]{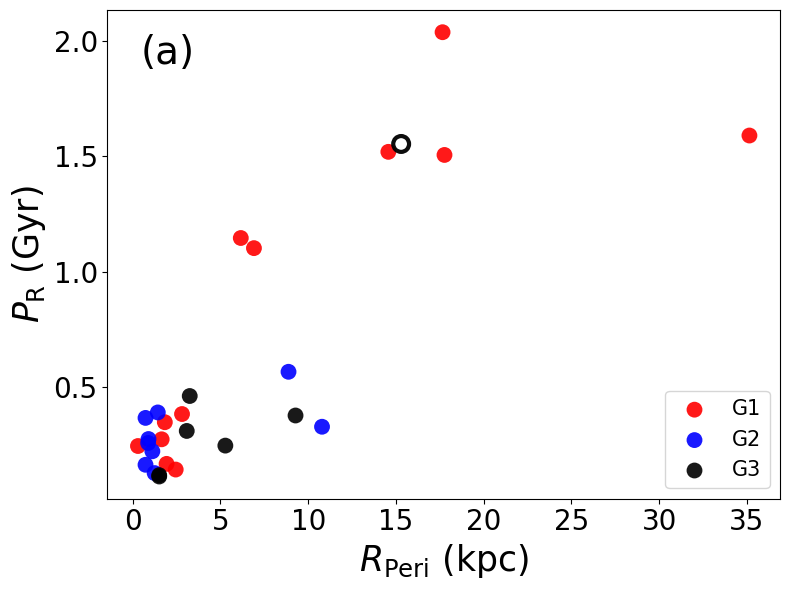}
\includegraphics[width=0.47\textwidth]{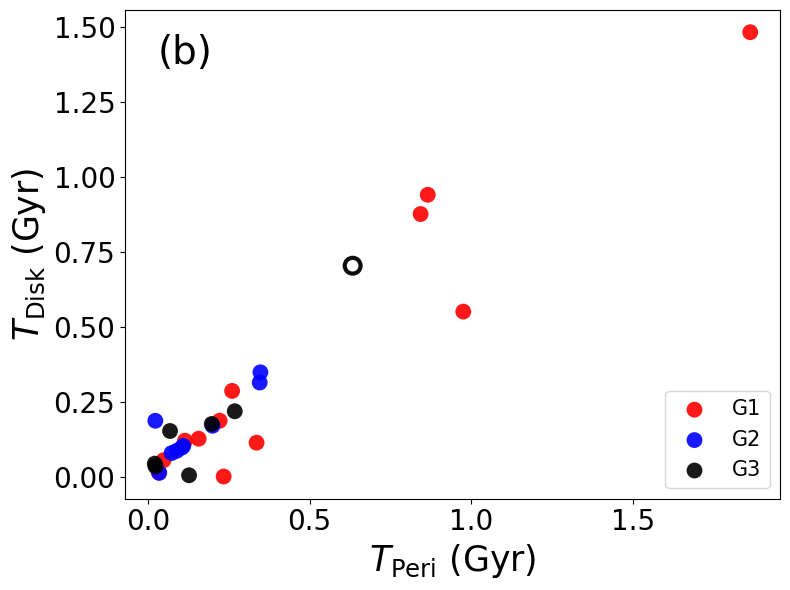}

\center
\caption{Orbital parameter relationships for clusters in different morphological groups. (a) The distribution of pericentric distance ($R_{\rm Peri}$) versus radial orbital period ($P_{\rm R}$). (b) The time since the last pericentric passage ($T_{\rm Peri}$) versus the time since the most recent disk crossing ($T_{\rm Disk}$). Clusters are color-coded by morphological group: G1 in red, G2 in blue, and G3 in black. Pal 12 is highlighted with an open marker.}
\label{para3}
\end{figure}

\textit{$R_{\rm Peri}$} and \textit{$P_{\rm R}$}: Panel (a) of Fig. \ref{para3} shows the distribution of our sample clusters as a function of $R_{\rm Peri}$ and $P_{\rm R}$. The parameter $R_{\rm Peri}$ marks the region where stars are most susceptible to tidal stripping and where the Galactic tidal field is strongest, while $P_{\rm R}$ reflects how frequently a cluster approaches the Galactic center. These two parameters together therefore provide a measure of the strength and frequency of tidal interactions. A roughly positive correlation is seen between $R_{\rm Peri}$ and $P_{\rm R}$, with many clusters concentrated in the lower-left region of the diagram. However, clusters exhibiting prominent tidal structures (G1) are distributed across a wide range of these parameters. Some G1 clusters even have relatively large $R_{\rm Peri}$ and long $P_{\rm R}$, suggesting that strong tidal features can still develop in systems that do not reside in the innermost regions of the Galaxy. This may also be related to the clusters' current positions and directions of motion along their orbital paths between pericenter and apocenter.

\textit{$T_{\rm Peri}$} and \textit{$T_{\rm Disk}$}: Panel (b) of Fig. \ref{para3} shows the relationship between the time since the last pericentric passage ($T_{\rm Peri}$) and the time since the most recent disk crossing ($T_{\rm Disk}$). Both parameters serve as indicators of whether a cluster has recently experienced strong tidal interactions. As shown in the figure, a positive correlation is also evident between the two, as clusters with shorter orbital periods tend to undergo pericentric passages and disk crossings more frequently. Most clusters are concentrated at small $T_{\rm Peri}$ and $T_{\rm Disk}$ values, suggesting that they have experienced significant tidal perturbations in the recent past. Notably, several G1 clusters are located toward the upper-right part of the diagram, implying that their last strong tidal interactions occurred longer ago, yet they still exhibit prominent tidal structures. This indicates that tidal features can persist for a considerable time after a major interaction, and their visibility is not solely determined by the time elapsed since the last pericentric or disk-crossing event but may also be constrained by the clusters' internal dynamical properties.
    
\begin{figure}
\center

\includegraphics[width=0.48\textwidth]{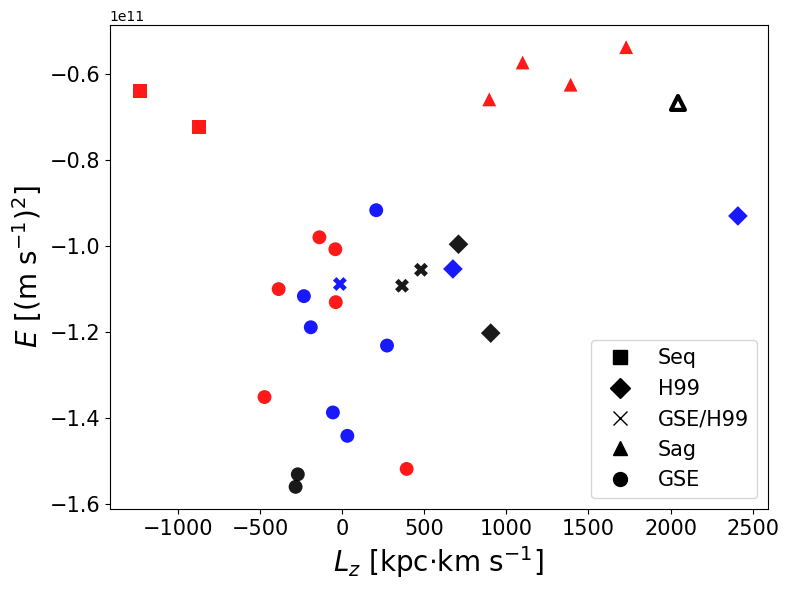}

\center
\caption{Distribution of GCs in the $L_{z}$-$E$ space.
Symbols indicate possible progenitor origins: circles represent clusters likely associated with the Gaia-Sausage-Enceladus (GSE) merger, triangles denote clusters linked to the Sagittarius (Sag) dwarf galaxy, squares indicate possible origin in the Sequoia (Seq) event, diamonds denote clusters associated with the Helmi stream (H99), and crosses mark clusters that may belong to either Gaia-Sausage-Enceladus or the Helmi stream (GSE/H99). Clusters are color-coded according to morphological classification: red for G1, blue for G2, and black for G3. Pal 12 is highlighted with open marker.}
\label{para4}
\end{figure}

\textit{Progenitor}: To assess whether accretion origin plays a role in shaping the outer structures of GCs, we examined their distribution in the energy-angular momentum ($E$-$L_z$) plane (Fig. \ref{para4}), a diagnostic that has been widely applied to trace the assembly history of the Galactic halo and to connect clusters with past merger events \citep[see, e.g.,][]{2019A&A...630L...4M, 2020ApJ...901...48N, 2022ApJ...926..107M}. The associations between clusters and their progenitors adopted in this study are based on \citeauthor{2019A&A...630L...4M} (\citeyear{2019A&A...630L...4M}, \citeyear{2025RNAAS...9...64M} update) and \citet{2020MNRAS.493..847F}.

As expected, clusters associated with major accretion events occupy relatively distinct regions in this space: Sagittarius members are characterized by high angular momenta and relatively unbound orbital energies; Sequoia clusters stand out for their strongly retrograde motions (i.e., large negative $L_z$); and Gaia-Sausage-Enceladus clusters span a broader range in both $E$ and $L_z$, reflecting their dynamically heated, eccentric orbits following the merger. From a morphological perspective, among the 14 Gaia-Sausage-Enceladus clusters, 6 are classified as G1, 6 as G2, and 2 as G3. Of the 5 Sagittarius clusters, 4 fall into G1 and 1 into G3. The 4 clusters associated with the Helmi stream are distributed as 2 G2 and 2 G3, while both Sequoia clusters are classified as G1. In addition, there are 3 clusters of uncertain origin (possibly linked to either Gaia-Sausage-Enceladus or the Helmi stream) including 1 G2 and 2 G3. Overall, the distribution of morphological types across different accretion groups shows certain systematic differences. In particular, clusters associated with Sagittarius and Sequoia appear to exhibit a closer correspondence between tidal morphology and their accretion origin. However, this trend remains tentative due to the limited sample size, and thus cannot be regarded as definitive evidence that accretion history alone drives the formation of extended tidal structures.

Taken together, our analysis indicates that clusters with tidal features are distributed across a wide range of pericentric distances, orbital periods, pericentric passage times, and disk-crossing times, showing no clear preference for regions experiencing stronger or more frequent tidal forces. Likewise, accretion history provides no definitive evidence of driving the formation of extended tidal structures.

\subsection{Limitations in the detection of tidal structures}

Despite recent advances in wide-field surveys and data processing techniques, the identification of tidal structures around Galactic GCs remains subject to several observational and methodological limitations. These factors may introduce biases in the detection and interpretation of extended features, potentially leading to an underestimation of their prevalence, morphology, or physical origin.

\paragraph{Limitations of data depth}

The data depth is crucial for assessing the completeness of tidal structures. In this
work, we make use of imaging data from the DESI Legacy Survey, which reaches down to $r$ or $g$ $\sim$23-24 mag. For the GCs in our sample, this depth corresponds to 2-3 magnitudes below the main-sequence turnoff, thereby encompassing the majority of stellar populations associated with the clusters. Importantly, the Legacy Survey represents the deepest publicly available wide-field dataset that contains the largest number of GCs. Thus, the data employed here greatly reduces the biases arising from limited photometric depth, allowing for a more complete characterization of tidal features.

In addition, for each cluster, we compared our matched-filter results with previous studies, including those based on deeper imaging (see the cluster-specific discussions in Appendix \ref{sec:appendix_clusters}), in order to evaluate the consistency of the detected features. This cross-validation approach helps minimize the risk of misclassification due to depth-related biases and ensures the reliability of our morphological classification.

\paragraph{Impact of binning and smoothing scales}

Methods such as matched filtering \citep{2002AJ....124..349R} require careful choices of the spatial binning scale and the size of the smoothing kernel applied to density maps. Excessively large bins or overly broad kernels can make thin tidal tails difficult to discern, whereas overly small bins and kernels may introduce excessive noise and generate spurious structures. Therefore, by testing a range of spatial binning and smoothing scales and cross-checking our results against independent studies (see Appendix \ref{sec:appendix_clusters}), we minimize the risk of misclassification arising from parameter choices and ensure the reliability of our morphological classification.

\paragraph{Choice of tidal radius}

The tidal boundary of a globular cluster is often approximated by the King model tidal radius  \citep{1962AJ.....67..471K}, derived from fitting the surface density profile under the assumption of dynamical equilibrium. However, this estimate does not take into account orbital effects or the time-varying nature of the Galactic potential. Dynamical alternatives (such as the Jacobi radius and, in this work, the dynamical tidal radius at the current orbital phase) depend on the cluster's mass, orbital phase, and the adopted Galactic potential model. Using different $r_{\rm t}$ values can change the classification of outer stars as "bound" or "extratidal", thereby affecting whether a cluster is deemed to possess extended structures.

To mitigate these effects, we adopted both the widely used $r_{\rm t}^{\rm King}$ (\citeauthor{1996AJ....112.1487H} \citeyear{1996AJ....112.1487H}, 2010 edition) from the literature and $r_{\rm t}^{\rm dyn}$ (from the globular cluster databases referenced in this work) calculated from orbital parameters. Both sets of values are taken from well-established and authoritative cluster catalogs, ensuring consistency with previous studies. In our morphological classification, we explicitly considered both definitions, thereby reducing the limitations associated with relying on a single tidal radius and enhancing the reliability of extratidal structure identification.

\paragraph{Projection effects and viewing angles}

Tidal structures are inherently three-dimensional and typically align along a cluster's orbital path. When viewed edge-on, these features are prominently detectable; however, if the tail lies along the line of sight, projection may smear the signal and obscure its geometry. The likelihood of detecting such structures thus strongly depends on the orbital inclination relative to the observer. Moreover, since tail morphology can evolve rapidly with orbital phase, snapshots may capture the cluster at a time when its features are faint, asymmetric, or highly dispersed. We also examined the distributions of radial velocities and proper motions for the G2 and G3 clusters in three-dimensional space; unfortunately, we did not find any significant or systematic trends.

In summary, although matched filtering and wide-field imaging have improved our ability to identify extended structures in GCs, the detection remains vulnerable to various systematic effects. In this study, we minimized these effects by employing a homogeneous dataset, adopting uniform filtering parameters and processing procedures, and cross-validating our results with previous imaging studies. These measures reduce potential systematic biases and enhance the reliability of the identified tidal features and their morphological classifications.

\section{Summary}

In this study, we conducted a systematic investigation of the extratidal structures associated with Galactic GCs of likely extragalactic origin, aiming to explore their formation mechanisms and the key physical factors driving their morphological diversity. Using a homogeneous dataset from the DESI Legacy Imaging Surveys and a uniform matched-filtering procedure, we identified and morphologically classified the tidal features of our cluster sample. We then explored the connections between extratidal structures and clusters' intrinsic properties, orbital dynamical parameters, and Galactic environment, followed by a discussion of several observational and methodological limitations affecting their detection and classification. Within this framework, we arrived at the following results:

1. Our analysis shows that a substantial fraction of clusters exhibit extended tidal structures, including prominent tidal tails (G1) and diffuse stellar envelopes (G2), which in many cases extend well beyond both the King tidal radius and the dynamical tidal radius. In contrast, some clusters remain spatially confined within their nominal tidal boundaries and show no evidence of extratidal features (G3). In total, 12 clusters are classified as G1, 9 as G2, and 7 as G3.

2. By examining the intrinsic parameters of the clusters, we identified several clear trends. For example, G1 clusters typically exhibit lower total mass, lower escape velocity ($v_{\rm esc}$), lower concentration ($c$), and higher degree of tidal filling ($r_{\rm h,m}/r_{\rm t}^{\rm dyn}$), with statistically significant differences across morphological types. These patterns highlight the role of the gravitational potential well in suppressing stellar escape, while also emphasizing the internal fragility that makes clusters more prone to tidal disruption, even though a certain degree of overlap remains in both one- and two-dimensional parameter spaces.

3. In the dynamical context, our analysis reveals weak trends in tidal morphology and identifies $R_{\rm Peri}$, $e$, and $\theta_{\rm R}$ as potentially group-sensitive parameters. Clusters with smaller pericentric distances, higher eccentricities, and specific ranges of $\theta_{\rm R}$ are more likely to exhibit higher mass-loss fractions, consistent with the scenario of enhanced stripping near pericenter and increased mass loss during repeated disk crossings. However, substantial overlap remains among different morphological types. Comparisons between the detected tidal structures, the orbital path, and the direction toward the Galactic center show that although some clusters display alignments consistent with orbital stripping or Galactic tides, many exhibit more complex or nearly symmetric envelope-like morphologies with no clear preferred orientation. This diversity suggests that the formation of tidal structures may be more complex than previously expected.

4. By evaluating the potential influence of the Galactic environment through some parameters such as pericentric distance ($R_{\rm Peri}$), radial orbital period ($P_{\rm R}$), time since the last pericentric passage ($T_{\rm Peri}$), and time since the most recent disk crossing ($T_{\rm Disk}$), we find that clusters with tidal features are distributed across a wide range of parameter space. This indicates that neither stronger tidal fields near the Galactic center nor more frequent tidal interactions necessarily drive the formation of outer structures. Moreover, the presence of extended features among clusters of different accretion origins suggests that accretion history itself has not been a dominant factor in producing tidal structures.

5. We discussed several observational and methodological limitations that may affect the identification of tidal structures, including data depth, smoothing scales, and the choice of tidal boundary. In this study, we mitigated these issues by adopting a homogeneous dataset from the DESI Legacy Imaging Surveys, applying a uniform matched-filtering and smoothing scheme, cross-validating our results against those from the literature, and explicitly considering both the King tidal radius and the dynamical tidal radius from authoritative globular cluster catalogs in our morphological classification. These measures effectively reduce systematic biases and enhance the reliability of tidal structure identification.

Overall, our study supports the view that the formation of extratidal structures is not governed by a single factor but arises from the complex interplay between a cluster’s internal properties and the external Galactic environment. Internal parameters, particularly those linked to the depth of the gravitational potential well, determine a cluster’s fundamental susceptibility to stellar stripping, while the external environment largely regulates the efficiency of this process. Together, these coupled effects shape both the likelihood of developing extratidal features and the morphological diversity observed among Galactic GCs.

\begin{acknowledgements}

Jundan Nie acknowledges the support of the National Key R\&D Program of China (grant Nos. 2021YFA1600401 and 2021YFA1600400), the National Natural Science Foundation of China (NSFC) (grant No. 12373019), the Beijing Natural Science Foundation (grant No. 1232032), and the science research grants from the China Manned Space Project (grant Nos. CMS-CSST-2021-B03, CMS-CSST-2021-A10 and CMS-CSST-2025-A11). Biwei Jiang acknowledges the support of the National Natural Science Foundation of China (NSFC) (grant No. 12133002). Hao Tian acknowledges the support of the National Key R\&D Program of 
China (No. 2024YFA1611902).
\end{acknowledgements}

\bibliographystyle{aa}
\bibliography{wangszGC}

\begin{appendix}

\section{Detailed notes on individual clusters}
\label{sec:appendix_clusters}

In this section, we provide detailed notes on the classification of each cluster, including comparisons with previous literature and our matched-filter analysis.

\textbf{\noindent(1) NGC 288:} This cluster is classified as G1 based on the presence of tail-like structures extending beyond $r_{\rm t}^\mathrm{dyn}$. This result is consistent with previous detections of tidal features in deep DECam imaging from the Dark Energy Survey \citep{2018ApJ...862..114S} and Gaia EDR3 data \citep{2021ApJ...914..123I}, and also agrees with the classification of \citet[][hereafter PCB20]{2020A&A...637L...2P}.

\textbf{\noindent(2) NGC 362:} \citet{2019MNRAS.486.1667C} reported the presence of short tidal tails around this cluster based on Gaia DR2 data, and accordingly, PCB20 classified it as a G1 cluster. However, our matched-filter S/N map only reveals structures above the 3$\sigma$ level extending beyond $r_{\rm t}^\mathrm{King}$, with no features detected beyond $r_{\rm t}^\mathrm{dyn}$. We attribute the discrepancies to differences in methodological applicability. In our matched-filter analysis, the weak tidal signal is overwhelmed by the extremely high stellar density of the Small Magellanic Cloud when computing the signal-to-noise ratio. Therefore, given the classification scheme adopted in this paper and our requirement for uniformity in data and methodology, we rely on our own detection and classify this cluster as a G2-type system.

\textbf{\noindent(3) Whiting 1:} This cluster is likely an accreted member of the Sagittarius dwarf galaxy \citep{2010ApJ...718.1128L, 2019A&A...630L...4M}. \citet{2017MNRAS.467L..91C}, using MegaCam imaging from both CFHT and Magellan, and \citet{2022ApJ...930...23N}, based on DECam photometry from the DECaLS survey, both reported structures extending beyond the tidal radius through matched-filter analysis. In contrast, \citet{2018ApJ...860...66M} (also using CFHT/Magellan MegaCam data) and \citet{2022MNRAS.513.3136Z} (using deep DECam imaging) found no significant extratidal features. Our matched-filter S/N map reveals a tail-like feature extending beyond $r_{\rm t}^\mathrm{dyn}$ with an elongated morphology, consistent with the former results. This discrepancy may stem from the higher sensitivity of the matched-filter technique to low-surface-brightness structures. Based on these characteristics, we classify it as a G1 cluster.

\textbf{\noindent(4) NGC 1261:} Some indications of tail-like features beyond $r_{\rm t}^\mathrm{dyn}$ are present in our matched-filter map, which are further confirmed by the angular density distribution (tail-like features from the lower left to the upper right). This is broadly consistent with previous studies. Based on deep DECam photometry, \citet{2018MNRAS.473.2881K} reported a symmetric power-law envelope, and thus PCB20 classified the cluster as G2. However, deeper DES photometry by \citet{2018ApJ...862..114S} and Gaia EDR3 data from \citet{2021ApJ...914..123I} revealed more extended extratidal structures. More recently, \citet{2025A&A...693A..69A} used data from the Southern Stellar Stream Spectroscopic Survey (S$^5$) and Gaia DR3, identifying a large number of high-probability extratidal member stars beyond $r_{\rm J}$ of NGC 1261. Considering these results along with our own, we classify NGC 1261 as a G1 cluster.

\textbf{\noindent(5) NGC 1851:} Previous studies have reported that this cluster hosts an extended envelope reaching roughly $2^\circ$ from its center \citep{2009AJ....138.1570O,2014MNRAS.445.2971C,2018MNRAS.474..683C,2018MNRAS.473.2881K}.
An analysis based on Gaia DR2 \citep{2020MNRAS.495.2222S} did not reveal additional large-scale features, whereas Gaia EDR3 data \citep{2021ApJ...914..123I} suggest the presence of diffuse tail-like structures extending nearly $10^\circ$ across the sky.
In our matched-filter S/N map, we detect features extending beyond $r_{\rm t}^\mathrm{dyn}$, but their appearance is more consistent with a broad, diffuse envelope rather than narrow tidal tails.
Given the observed morphology and the depth of our data, we classify this cluster as G2.

\textbf{\noindent(6) NGC 1904:} \citet{2018MNRAS.474..683C} and \citet{2022MNRAS.513.3136Z} reported the detection of a symmetric power-law envelope around this cluster based on deep DECam photometry. However, our matched-filter S/N map reveals clear tidal tails extending well beyond $r_{\rm t}^\mathrm{dyn}$. These features are elongated along the north-south direction and are consistent with the structures identified by \citet{2018ApJ...862..114S} using DES deep photometry. More recently, \citet{2025A&A...693A..69A} also used data from the Southern Stellar Stream Spectroscopic Survey (S$^5$) and Gaia DR3 to identify high-probability extratidal member stars beyond $r_{\rm J}$ of NGC 1904. We therefore classify NGC 1904 as a G1 cluster.

\textbf{\noindent(7) NGC 4147:} According to the study by \citet{2022A&A...665A...8K}, this cluster may be associated with the Sagittarius stream, and 11 extratidal member stars were identified, 6 of which lie beyond its Jacobi radius. Earlier, \citet{2010A&A...522A..71J} identified tidal tails around NGC 4147 using SDSS photometry, revealing a roughly north-south oriented two-arm morphology, although their data did not extend far beyond the main-sequence turn-off. More recently, \citet{2024AJ....168..237Z}, based on DESI photometric data, detected multiple-arm substructures outside the tidal radius. Similarly, our matched-filter results confirm the presence of comparable multiple-arm tidal tails. Considering all these findings, we classify NGC 4147 as a G1 cluster.

\textbf{\noindent(8) NGC 4590:}  Although PCB20 classified this cluster as G1 based on the detection of tidal stream from Gaia DR2 by \citet{2019MNRAS.488.1535P}, our matched-filter S/N map does not reveal any prominent tidal tail features. Instead, we identify a symmetric power-law envelope extending beyond $r_{\rm t}^\mathrm{King}$ and approaching $r_{\rm t}^\mathrm{dyn}$. Our finding is consistent with that of \citet{2020MNRAS.495.2222S}, who used Gaia data and detected a diffuse envelope surrounding the cluster. We find that the large differences in the reported tidal structures primarily result from the different methodologies adopted: study \citet{2019MNRAS.488.1535P} integrates the orbit and performs an all-sky search for cluster members, a strategy suited to identifying widely dispersed stellar streams, whereas both our analysis and \citet{2020MNRAS.495.2222S} focus on tidal features in the immediate vicinity of the cluster. Given this cluster-centric perspective and the consistency between our result and that of \citet{2020MNRAS.495.2222S}, we classify the cluster as a G2 cluster. 

\textbf{\noindent(9) NGC 5024:} Based on orbital calculations, \citet{2021ApJ...909L..26B} suggested that NGC 5024 may be associated with both the Sylgr stream \citep{2019ApJ...872..152I} and the Ravi stream \citep{2018ApJ...862..114S}. Previous photometric studies have reported no evidence of extratidal features around this cluster \citep{2010A&A...522A..71J, 2012MNRAS.419...14C, 2014MNRAS.445.2971C}.
A recent wide-field photometric study \citep{2025MNRAS.540.2863W} detected a mild stellar overdensity to the southwest of the cluster, spatially aligned with its orbital path and possibly representing weak tidal debris, but extending no farther than the dynamical tidal radius.
Our matched-filter S/N map likewise reveals no significant extratidal structure, in agreement with these findings.
Based on these results, we classify NGC 5024 as a G3 cluster, consistent with the classification adopted by PCB20.

\textbf{\noindent(10) NGC 5053:} Early SDSS-based studies \citep{2006ApJ...651L..33L, 2010A&A...522A..71J} reported tail-like structures, while \citet{2010AJ....139..606C}, based on wide-field deep photometry, suggested the presence of an envelope-like component surrounding the cluster and proposed the possible existence of a tidal bridge connecting it to the nearby NGC 5024. Subsequently, \citet{2019MNRAS.485.4906D} used Gaia DR2 data to suggest that the tidal structure of NGC 5053 may extend beyond its Jacobi radius. A recent wide-field photometric study \citep{2025MNRAS.540.2863W} detected a mild elongation of NGC 5053 along its western axis and an outer stellar overdensity aligned with its orbital path, but reported no clear evidence of a tidal bridge to NGC 5024. In contrast, some deeper photometric studies \citep{2012MNRAS.419...14C, 2014MNRAS.445.2971C} did not detect significant extratidal features, likely due to limited spatial coverage. From the morphology of our detected structure, the cluster shows a faint extension toward the southwest that extends beyond $r_{\rm t}^\mathrm{dyn}$. In addition, several diffuse, larger-scale features also lie outside $r_{\rm t}^\mathrm{dyn}$. Therefore, we classify this cluster as G2, consistent with the classification adopted by PCB20.

\textbf{\noindent(11) NGC 5272:} Based on orbital calculations, \citet{2021ApJ...909L..26B} suggested that NGC 5272 may be associated with the Svöl stream \citep{2019ApJ...872..152I}. Previous studies based on ground-based photometric data and Gaia observations \citep{2010A&A...522A..71J, 2014MNRAS.445.2971C, 2020MNRAS.495.2222S} have reported no evidence of extratidal structures around NGC 5272. A recent wide-field photometric analysis \citep{2025MNRAS.540.2863W} similarly found no clear extension beyond the tidal radius, although it noted the presence of faint stellar overdensities on both sides of the cluster along its orbital direction, which were interpreted as possible tidal tails. Our matched-filter S/N map does not reveal any significant extratidal structure, broadly consistent with the conclusions of these studies. Based on these results, we classify NGC 5272 as a G3 cluster, in agreement with the classification of PCB20.

\textbf{\noindent(12) NGC 5466:} This is a well-known globular cluster with prominent tidal tails, which have been extensively reported in previous studies based on wide-field photometry and stellar density analyses \citep[e.g.,][]{2006ApJ...639L..17G, 2010AJ....139..606C, 2010A&A...522A..71J}. Our matched-filter S/N map clearly reveals symmetric tidal features extending beyond $r_{\rm t}^\mathrm{dyn}$ on both sides of the cluster, in good agreement with the morphology reported in earlier studies. Based on this evidence, we classify NGC 5466 as a G1 cluster.

\textbf{\noindent(13) NGC 5634:} The study by \citet{2022A&A...665A...8K} suggested that this cluster may be associated with the Sagittarius stream, with 10 extratidal member stars identified, 5 of which lie beyond its Jacobi radius. Although \citet{2012MNRAS.419...14C,2014MNRAS.445.2971C} provided deep photometric data for NGC 5634, the limited spatial coverage of their observations restricted the evaluation of potential extratidal features. More recently, \citet{2025ApJ...988...39W} investigated the tidal structure of NGC 5634 using DESI data and found no significant extratidal features. Similarly, \citet{2025AJ....170..294C} reported no prominent tidal structures based on DELVE DR2 data. Based on the current observational evidence, we classify NGC 5634 as a G3 cluster.

\textbf{\noindent(14) NGC 5897:}
Studies \citet{2020MNRAS.495.2222S} and \citet{2025AJ....170..294C} both report extratidal features extending beyond the Jacobi radius, though with different morphologies: \citet{2020MNRAS.495.2222S} identifies two roughly symmetric extratidal features, whereas \citet{2025AJ....170..294C} finds only one prominent extension. In our analysis, we likewise detect extratidal structures well outside the Jacobi radius. The features we identify are bilateral, and even more pronounced than those reported in \citet{2020MNRAS.495.2222S}. Moreover, in the LSST 10-year deep data predictions presented in \citet{2025AJ....170..294C}, the two structures appear even more clearly. Based on the structural characteristics revealed in this work, together with the cluster classification criteria adopted in this paper, we conclude that the cluster is more consistent with a G1 type system.

\textbf{\noindent(15) NGC 5904:} Although previous studies (e.g., \citealt{2010A&A...522A..71J, 2019ApJ...884..174G, 2020MNRAS.495.2222S, 2021ApJ...914..123I}) have reported the presence of extended tidal tails around NGC 5904 based on SDSS and Gaia data, our matched-filter S/N map constructed from deep photometric data (reaching $r$-band depth of 23 mag) does not reveal any significant extratidal structures. The stellar density distribution appears nearly symmetric and is clearly confined within $r_{\rm t}^\mathrm{dyn}$, with no pronounced elongation along any specific direction. A recent wide-field photometric study \citep{2025MNRAS.540.2863W}, however, reported strong peripheral deformation, with tidal arm-like features extending north-south and symmetric overdensities along the east-west direction, and suggested that these structures are spatially connected to the cluster's known rotational axis. In our deeper data, we do not detect such features, indicating that they may possess extremely low surface brightness or be highly diffuse, thus falling below the detection threshold of our matched-filter analysis. Based on our results, we tentatively classify NGC 5904 as a G3 cluster.

\textbf{\noindent(16) NGC 6205:} Both \citet{2010A&A...522A..71J}, based on SDSS data analysis, and \citet{2020MNRAS.495.2222S}, using Gaia DR2 data, did not find any evidence of tidal tails or extratidal extensions around NGC 6205. Our matched-filter S/N map similarly shows no significant signs of tidal structures, in agreement with these previous studies. Therefore, we retain the classification of NGC 6205 as a G3 cluster, consistent with the classification by PCB20.

\textbf{\noindent(17) NGC 6229:} \citet{2018ApJ...860...66M}, based on deep MegaCam photometry, did not identify any significant extratidal features around NGC 6229. \citet{2019MNRAS.485.4906D} suggested that the cluster might exhibit some degree of extended structure within the theoretical Jacobi radius, based on the fitted density profile. Our matched-filter S/N map also reveals some extended features that clearly extend beyond $r_{\rm t}^\mathrm{King}$, though they remain mostly within $r_{\rm t}^\mathrm{dyn}$. These features may indicate the presence of marginal outer structures that have previously gone undetected. Based on these results, we conservatively classify NGC 6229 as a G2 cluster.

\textbf{\noindent(18) NGC 6341:} \citet{2020MNRAS.495.2222S} and \citet{2021ApJ...914..123I} identified a long tidal tail associated with NGC 6341 using Gaia data, and this feature was further confirmed by \citet{2020ApJ...902...89T} through deep CFHT photometry. A recent wide-field photometric study \citep{2025MNRAS.540.2863W} reported an elliptical extension of stellar density contours beyond the tidal radius, as well as a low-confidence extension aligned with the cluster's Galactocentric motion vector exhibiting a three-branched morphology, though its physical reality could not be firmly established due to its very low surface brightness. Our matched-filter S/N map also reveals tidal stretching, but the detected features remain confined within $r_{\rm t}^\mathrm{dyn}$ and do not extend to larger distances. This discrepancy may be related to the photometric depth limits of the DR10 dataset used in our analysis, as well as differences in the methodological implementations adopted across studies. Based on our observational results, we classify NGC 6341 as a G2 cluster.

\textbf{\noindent(19) Terzan 7:} This cluster has long been recognized as a member cluster of the Sagittarius dwarf galaxy \citep{2010ApJ...718.1128L, 2019A&A...630L...4M}. According to \citet{2022A&A...665A...8K}, a total of 491 extratidal member stars were identified in this cluster, 484 of which lie beyond its Jacobi radius. Our matched-filter S/N map reveals clear extended structures on both sides of the cluster, extending beyond $r_{\rm t}^\mathrm{dyn}$. The morphology of these features is consistent with typical tidal tails. Based on this evidence, we classify Terzan 7 as a G1 cluster.

\textbf{\noindent(20) Arp 2:} This cluster has also long been recognized as a member cluster of the Sagittarius dwarf galaxy \citep{2010ApJ...718.1128L, 2019A&A...630L...4M}. In the study of \citet{2022A&A...665A...8K}, 315 extratidal member stars were identified, including 271 located beyond the cluster's Jacobi radius. Our matched-filter S/N map reveals a marginal and asymmetric stellar extension, primarily toward the southwest of the cluster, extending beyond $r_{\rm t}^\mathrm{dyn}$. Based on this evidence, we classify Arp 2 as a G1 cluster.

\textbf{\noindent(21) Terzan 8:} This cluster is also a member of the Sagittarius dwarf galaxy \citep{2010ApJ...718.1128L, 2019A&A...630L...4M}.
In the study by \citet{2022A&A...665A...8K}, 110 extratidal member stars were identified in this cluster, 80 of which lie beyond its Jacobi radius. Our matched-filter S/N map reveals stellar extensions on both
the northern and southern sides of the cluster, which extend beyond $r_{\rm t}^\mathrm{dyn}$. Based on this evidence, we classify Terzan 8 as a G1 cluster.

\textbf{\noindent(22) NGC 6864:} \citet{2022MNRAS.513.3136Z} reported no significant extratidal structure around this cluster based on deep DECam data, while \citet{2022MNRAS.509.3709P} detected extended features beyond the tidal radius. In our results, we similarly find no clear evidence of tidal tails; however, a envelope-like structure is present beyond $r_{\rm t}^\mathrm{King}$ but still confined within $r_{\rm t}^\mathrm{dyn}$. Based on this evidence, we classify NGC 6864 as a G2 cluster.

\textbf{\noindent(23) NGC 6981:} Both \citet{2021A&A...646A.176P} and \citet{2022MNRAS.513.3136Z} suggest that NGC 6981 may possess a low-luminosity envelope. In our matched-filter S/N map, we do not detect tidal tail features, but we do identify an envelope-like structure that extends beyond $r_{\rm t}^\mathrm{dyn}$, consistent with the results of these studies. Based on these results, we classify NGC 6981 as a G2 cluster.

\textbf{\noindent(24) NGC 7089:} \citet{2010A&A...522A..71J}, based on SDSS data, did not find any evidence of extratidal structures around NGC 7089. However, \citet{2016MNRAS.461.3639K} identified a prominent power-law envelope based on deep DECam imaging, and \citet{2021ApJ...914..123I} detected extended tidal tails using Gaia EDR3 data. Our matched-filter S/N map does not reveal obvious tidal tail features, but a possible envelope-like structure is present. Our detection is in better agreement with the results from deeper DECam imaging, and we classify NGC 7089 as a G2 cluster..

\textbf{\noindent(25) NGC 7099:} \citet{2020MNRAS.495.2222S} reported the presence of prominent tidal tails based on Gaia DR2 data, whereas \citet{2020A&A...643A..15P} were unable to reproduce these features using deep DECam photometry, although their analysis revealed a short tidal tail and some scattered stellar debris. In our matched-filter S/N map, we do not detect any clear evidence of extended tidal tails, which may be due to the depth limitations of the DR10 photometric data. However, we do observe hints of peripheral elongation in the outer regions. Based on the current observational results, we classify NGC 7099 as a G3 cluster.

\textbf{\noindent(26) Pal 12:} This cluster is a known satellite of the Sagittarius dwarf galaxy \citep{2010ApJ...718.1128L, 2019A&A...630L...4M}. In the study by \citet{2022A&A...665A...8K}, 77 extratidal member stars were identified in this cluster, all of which lie beyond its Jacobi radius. However, our matched-filter S/N map shows no evidence of tail-like features associated with it. In addition, \citet{2018MNRAS.473.3062M} found no extended features based on deep photometry. It is worth noting that its tidal radius appear somewhat peculiar, with $r_{\rm t}^\mathrm{King}$ being much larger than $r_{\rm t}^\mathrm{dyn}$, and there seems to be no detailed discussion of this discrepancy in the existing literature. Therefore, we classify Pal 12 as a G3 cluster solely based on our matched-filter result, and this source is not included in the subsequent statistical analyses.

\textbf{\noindent(27) Pal 13:} \citet{2020AJ....160..244S} identified extended tidal tails associated with this cluster using DECaLS data, and \citet{2020A&A...635A..93P} also found that its member stars extend beyond the Jacobi radius. Our matched-filter S/N map clearly reveals the presence of tidal tail structures, which are aligned with the extended features reported by \citet{2020AJ....160..244S}, although their spatial extent is less pronounced in our data. Based on these results, we classify Pal 13 as a G1 cluster.

\textbf{\noindent(28) NGC 7492:} Our matched-filter S/N map reveals clear tidal tail structures, consistent with the findings of \citet{2017ApJ...841L..23N} based on Pan-STARRS PS1 data, and with the results of \citet{2025AJ....170..294C}. Although some studies using deeper photometric data \citep{2018ApJ...860...66M,2018ApJ...860...65M,2022MNRAS.513.3136Z} did not reproduce these features, we classify NGC 7492 as a G1 cluster based on our results and the aforementioned evidence of \citet{2017ApJ...841L..23N} and \citet{2025AJ....170..294C}.

\clearpage
\onecolumn
\section{Additional figures}

\begin{figure*}[htbp]
\center
\includegraphics[width=0.19\textwidth]{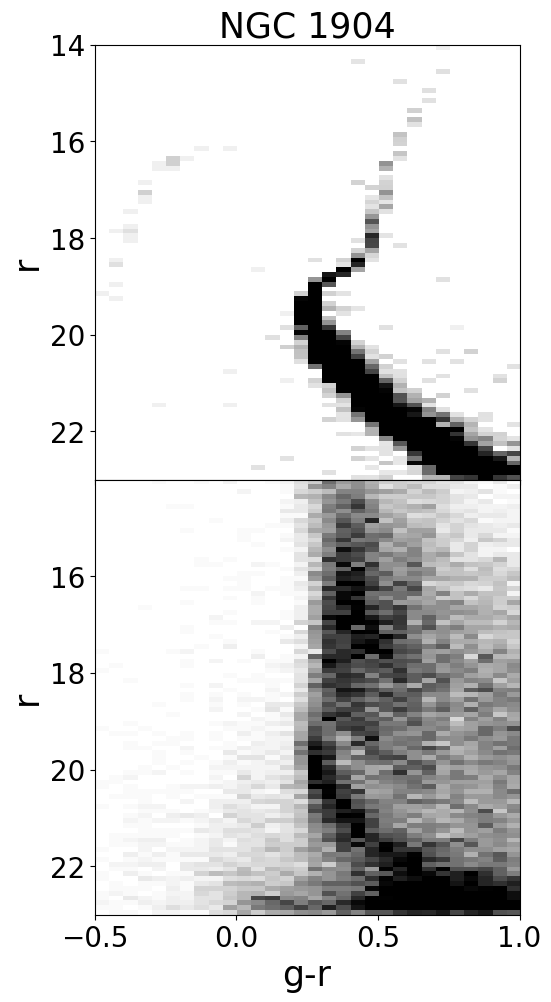}
\includegraphics[width=0.19\textwidth]{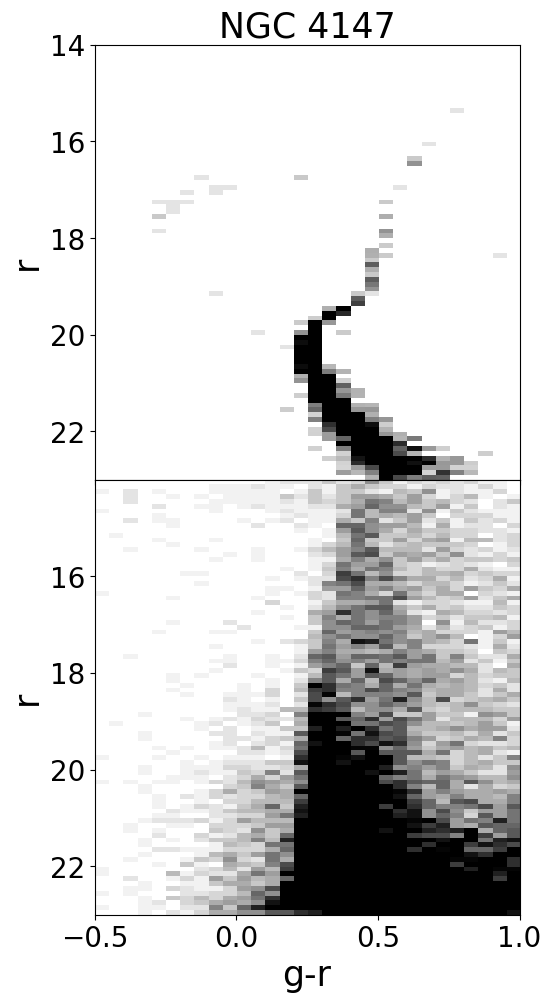}
\includegraphics[width=0.19\textwidth]{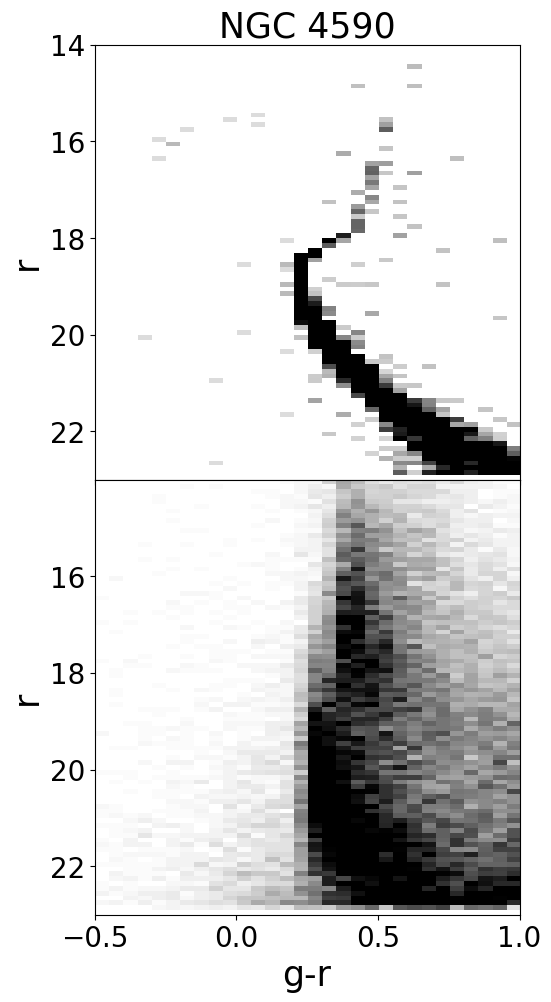}
\includegraphics[width=0.19\textwidth]{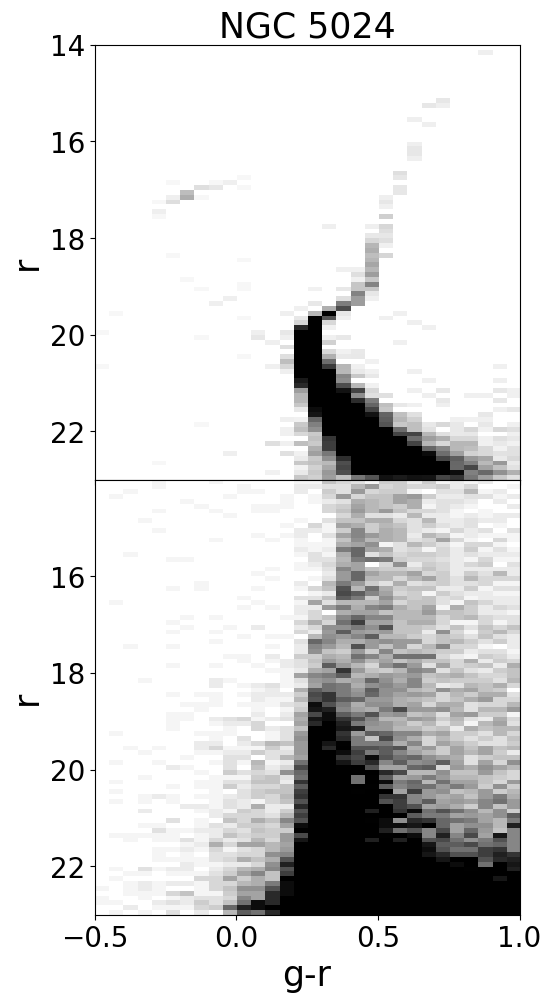}
\includegraphics[width=0.19\textwidth]{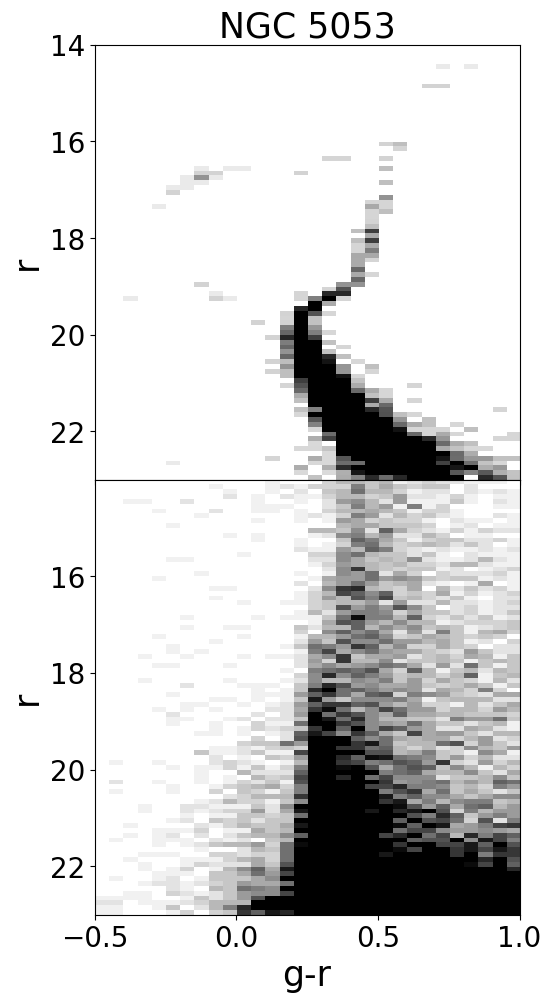}
\includegraphics[width=0.19\textwidth]{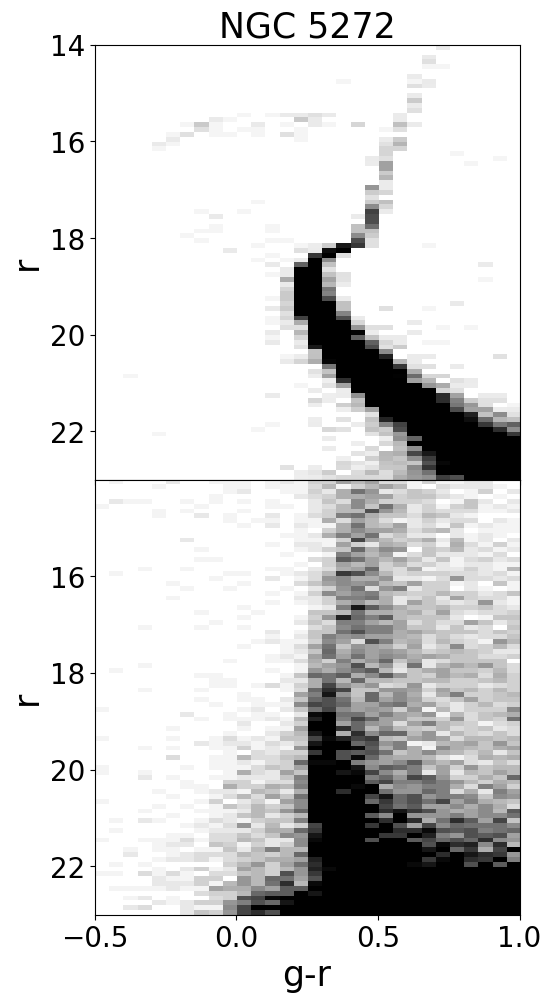}
\includegraphics[width=0.19\textwidth]{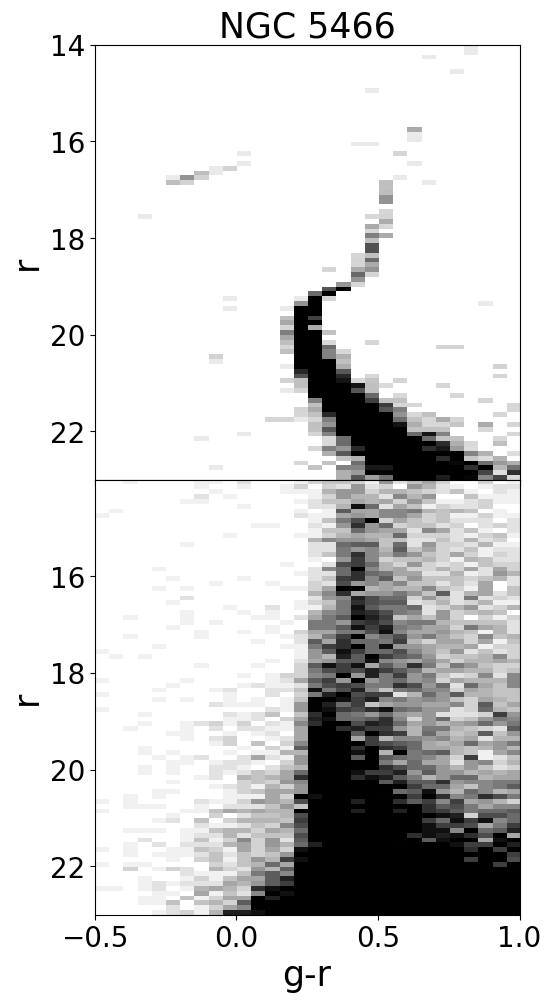}
\includegraphics[width=0.19\textwidth]{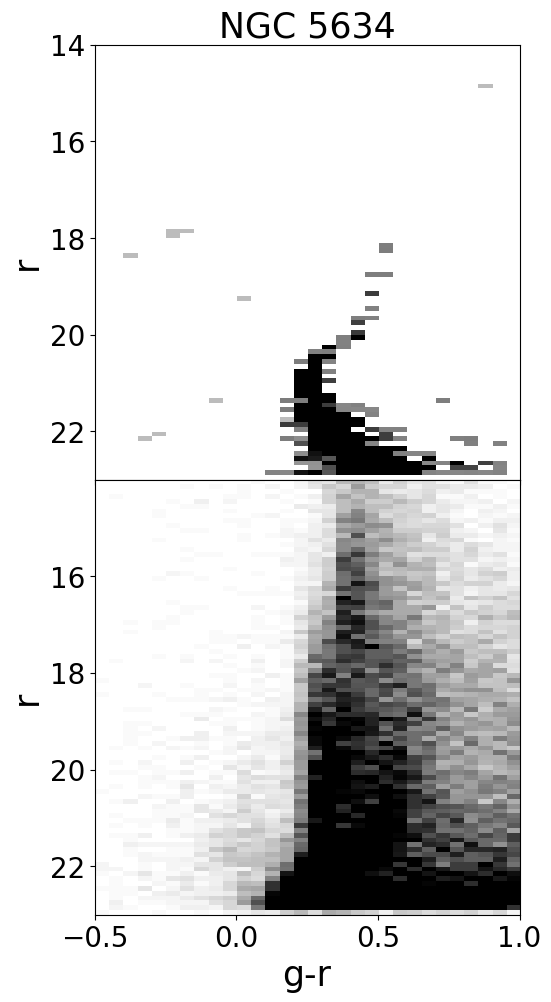}
\includegraphics[width=0.19\textwidth]{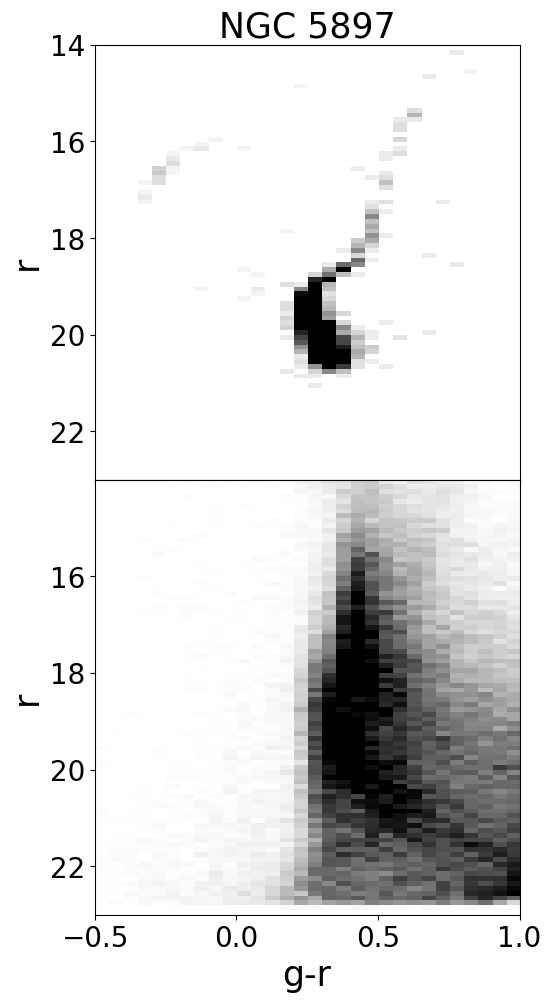}
\includegraphics[width=0.19\textwidth]{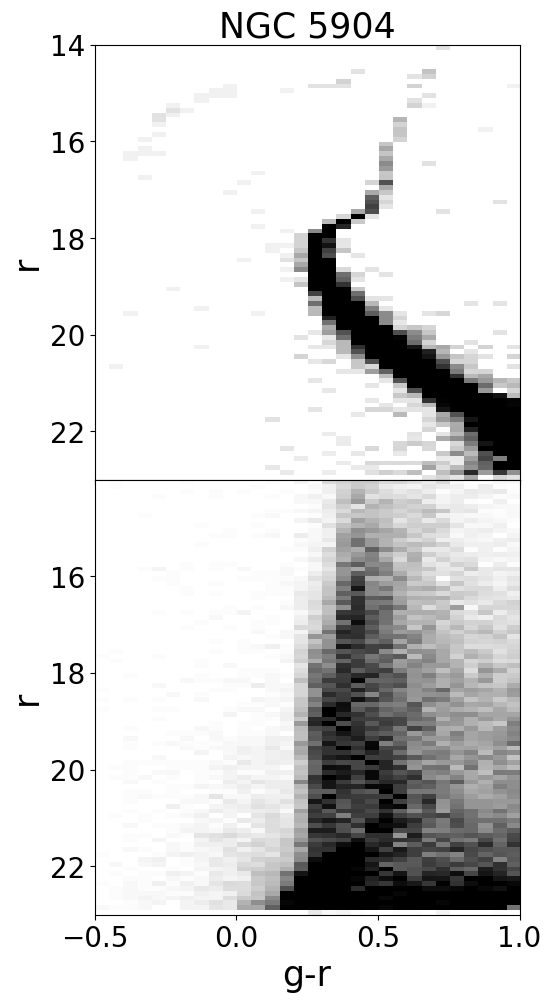}
\includegraphics[width=0.19\textwidth]{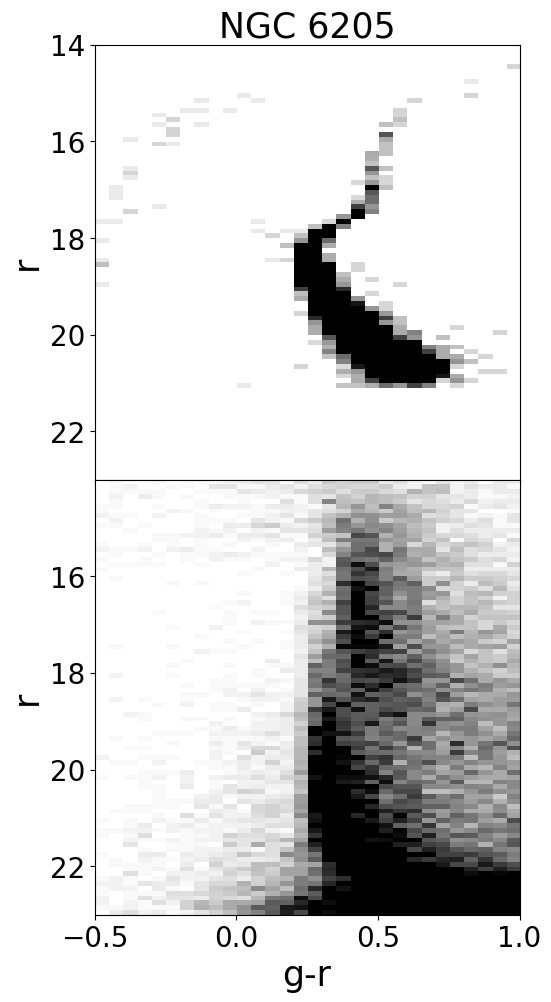}
\includegraphics[width=0.19\textwidth]{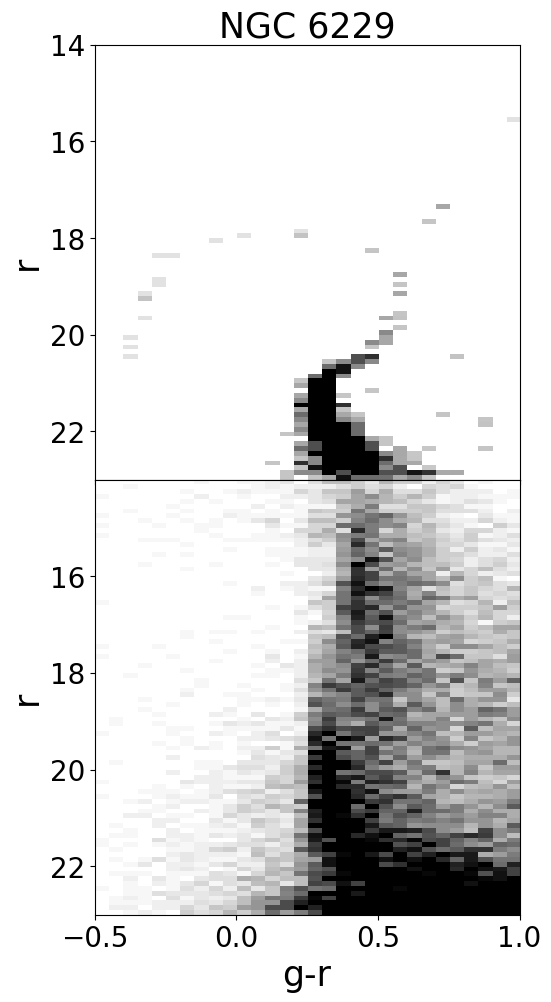}
\includegraphics[width=0.19\textwidth]{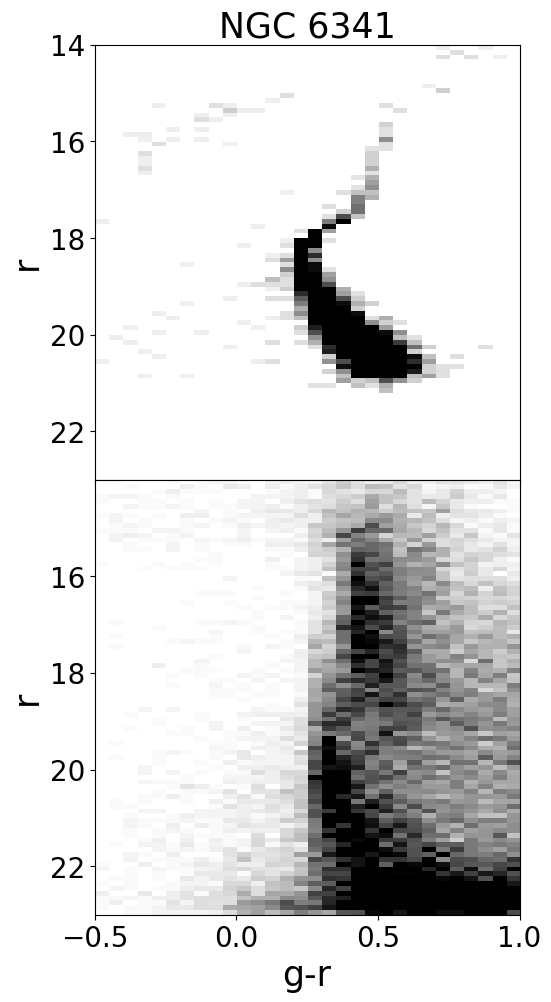}
\includegraphics[width=0.19\textwidth]{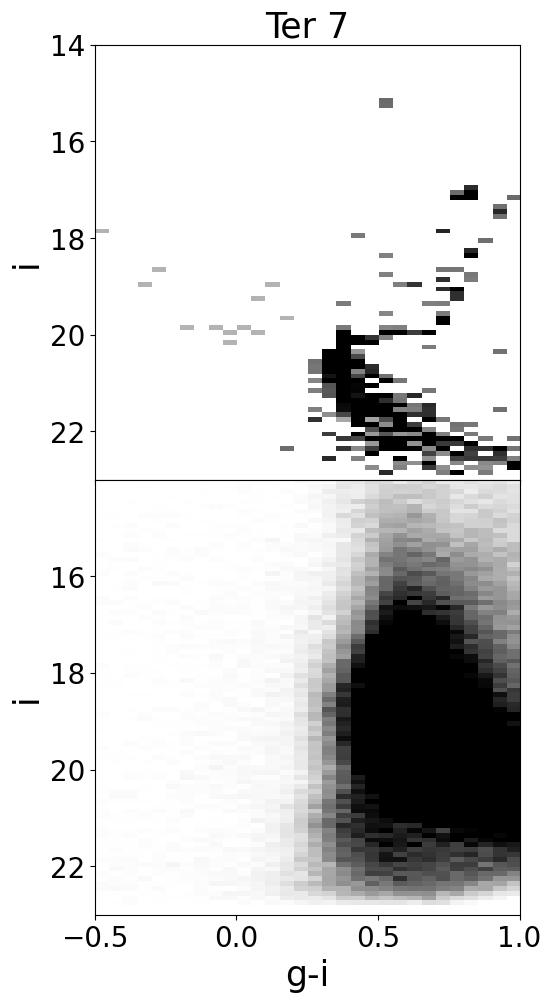}
\includegraphics[width=0.19\textwidth]{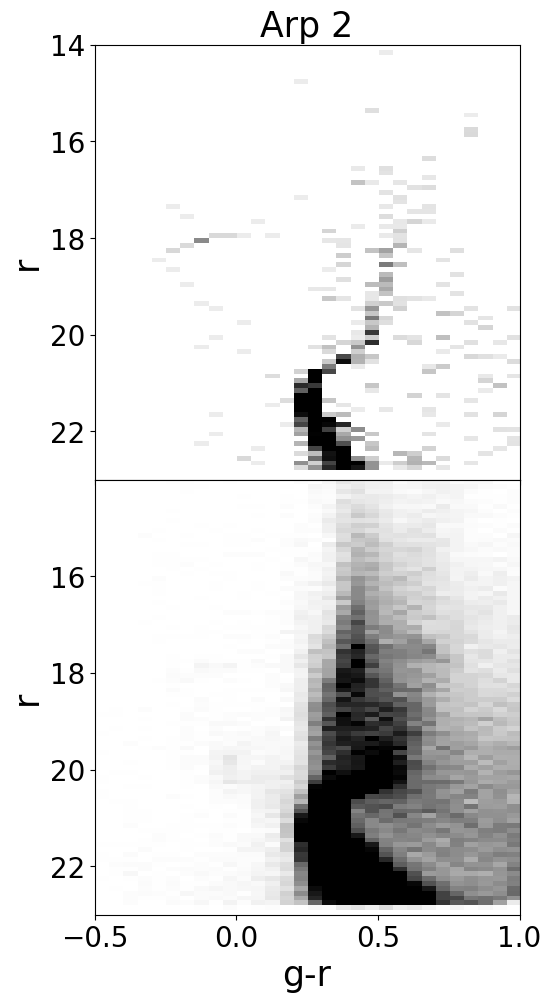}

\center
\caption{Color-magnitude diagrams for the remaining clusters in the sample, complementing those shown in Fig. \ref{CMD}}
\label{CMDappendix}
\end{figure*}

\begin{figure*}[htbp]
\ContinuedFloat
\center
\includegraphics[width=0.19\textwidth]{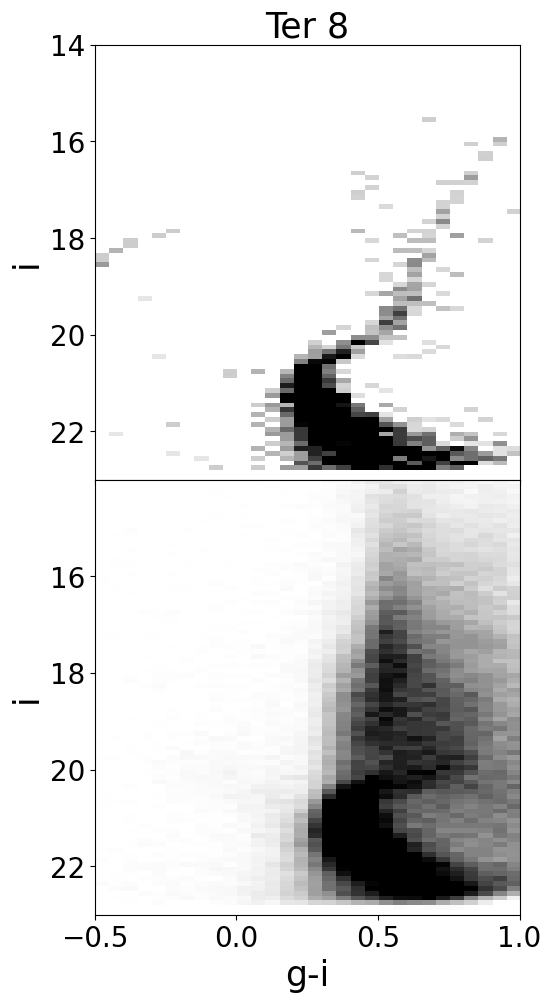}
\includegraphics[width=0.19\textwidth]{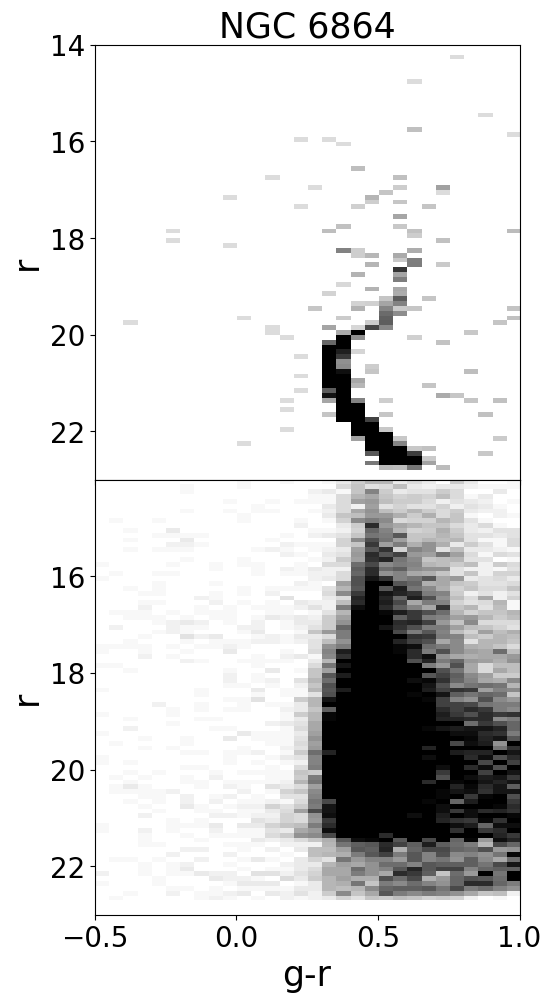}
\includegraphics[width=0.19\textwidth]{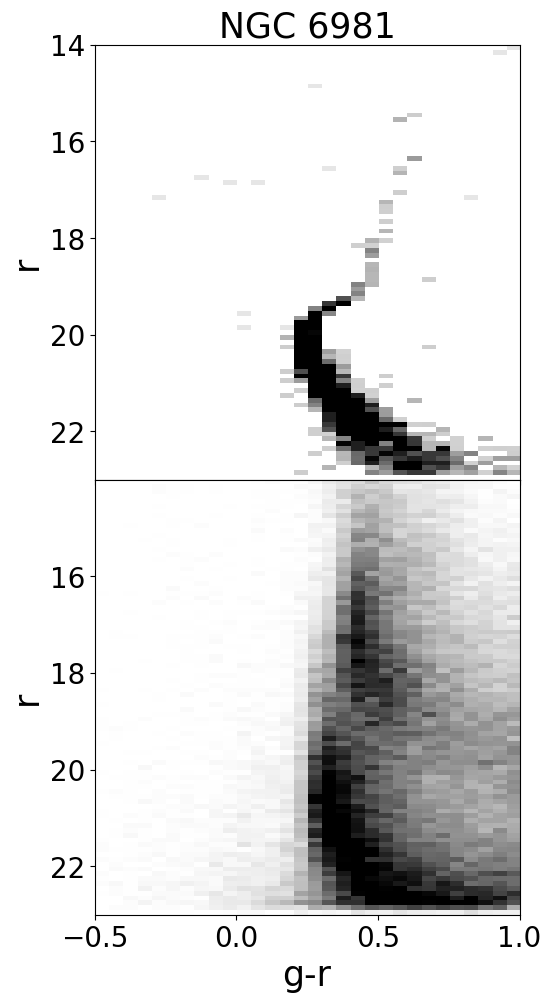}
\includegraphics[width=0.19\textwidth]{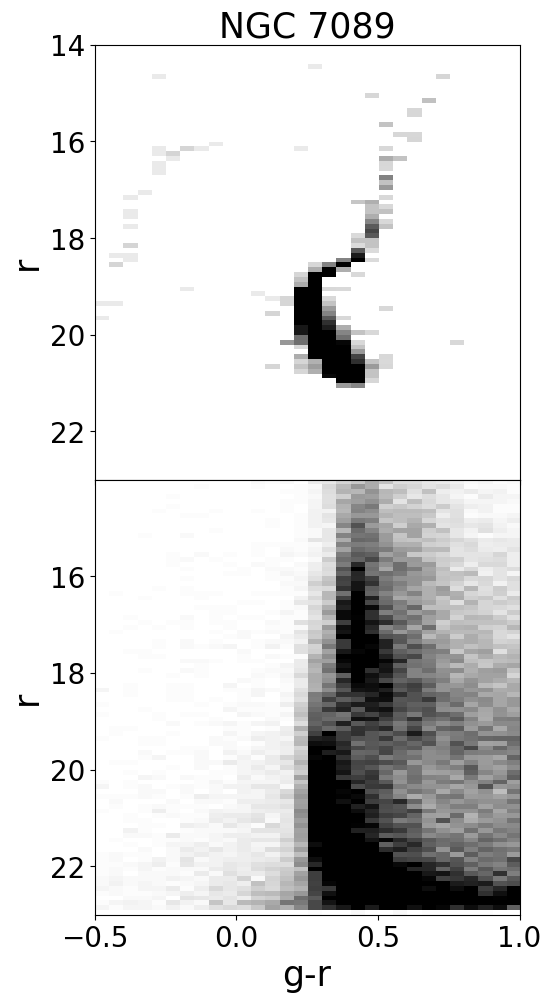}
\includegraphics[width=0.19\textwidth]{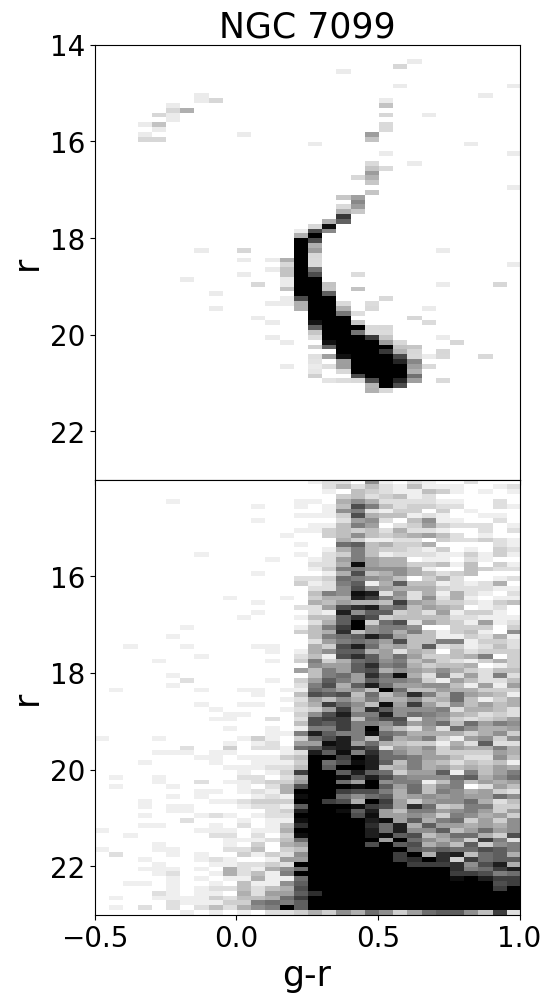}
\includegraphics[width=0.19\textwidth]{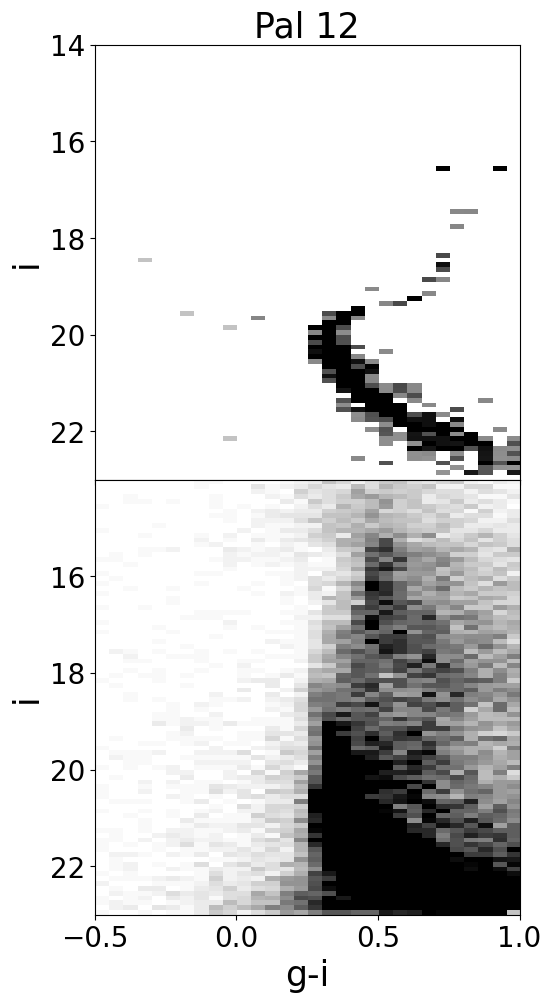}
\includegraphics[width=0.19\textwidth]{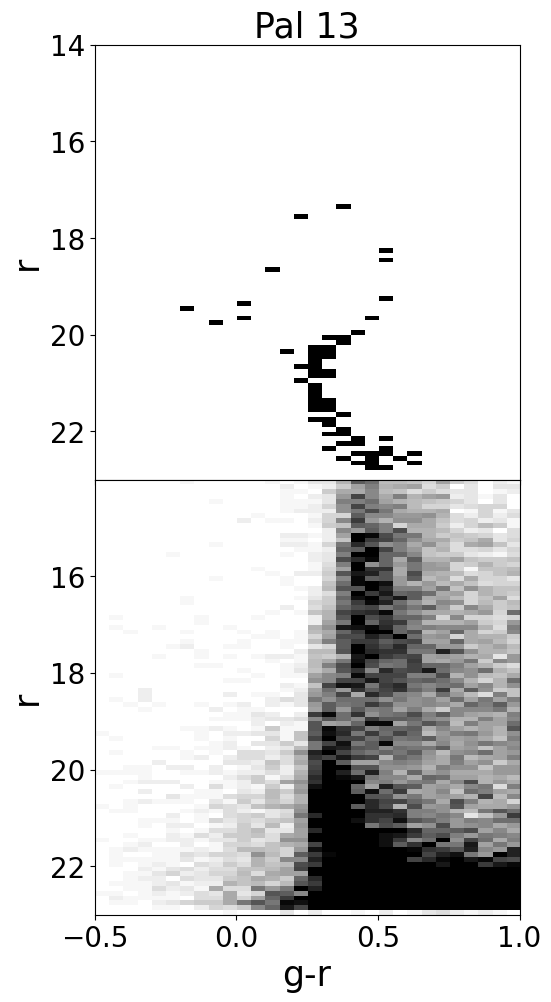}
\includegraphics[width=0.19\textwidth]{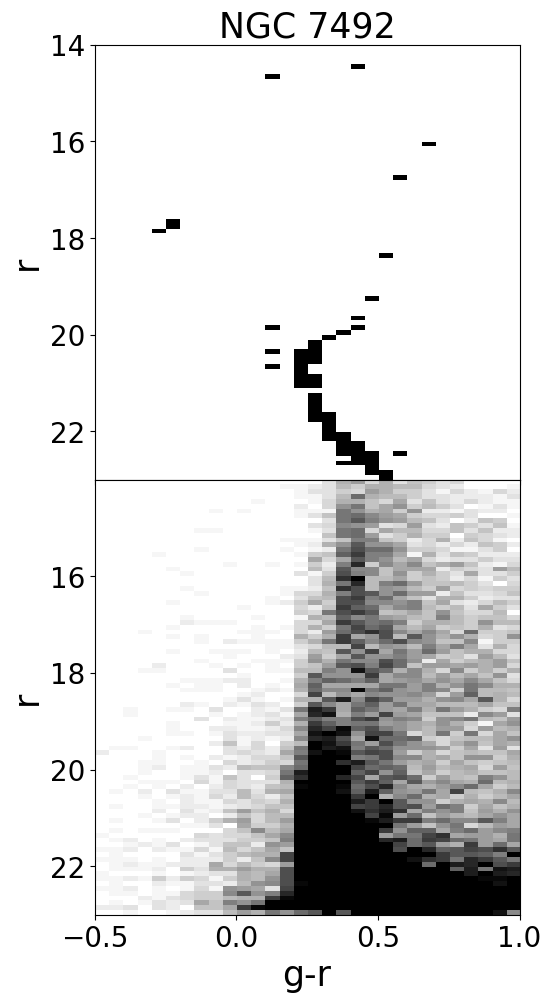}

\center
	\caption{Continued.}
\end{figure*}

\begin{figure*}[htbp]
\center
\includegraphics[width=0.32\textwidth]{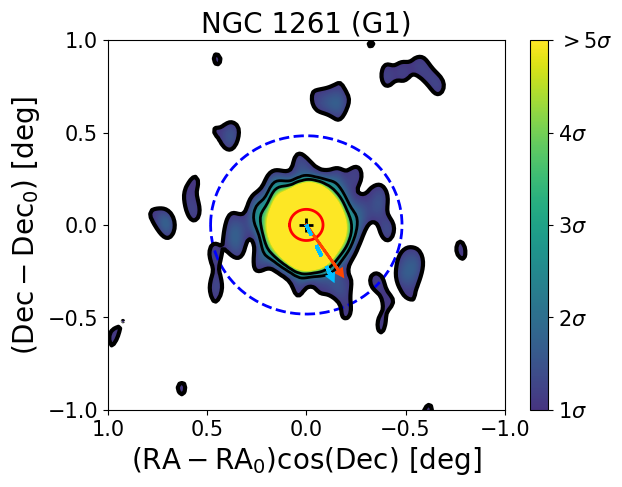}
\includegraphics[width=0.32\textwidth]{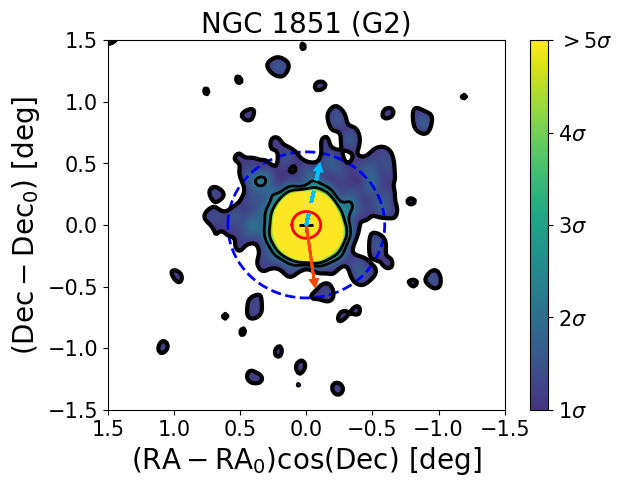}
\includegraphics[width=0.32\textwidth]{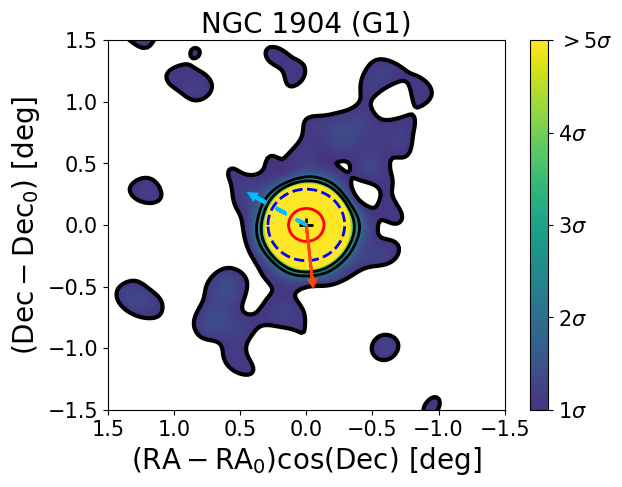}
\includegraphics[width=0.32\textwidth]{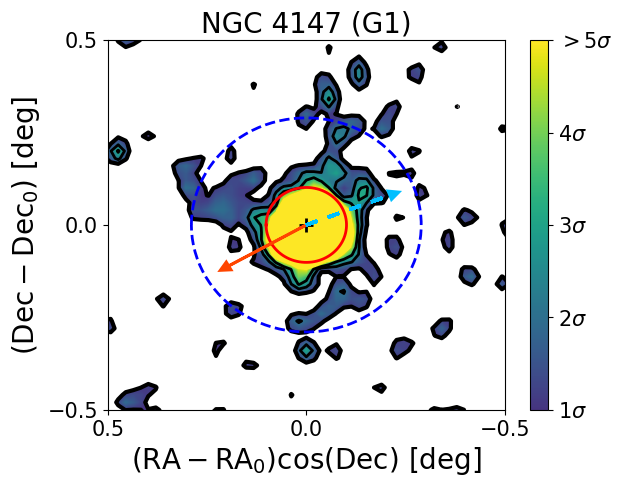}
\includegraphics[width=0.32\textwidth]{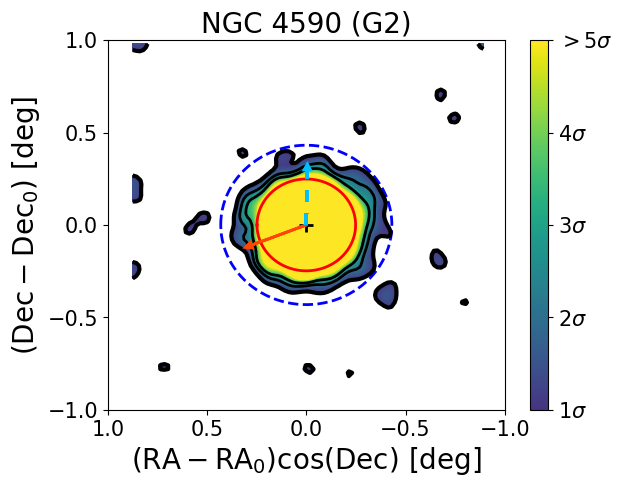}
\includegraphics[width=0.32\textwidth]{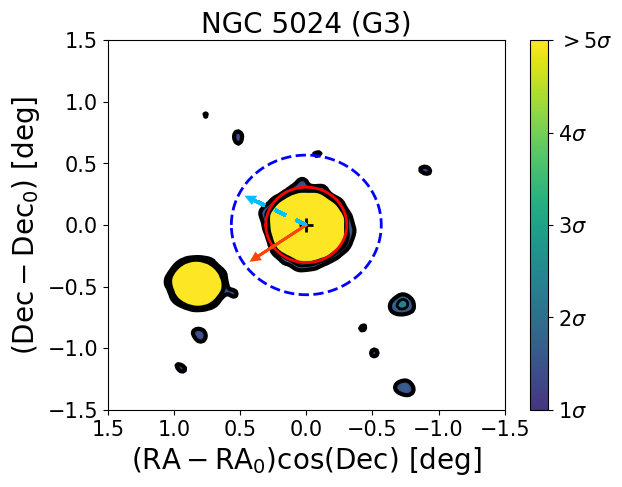}
\includegraphics[width=0.32\textwidth]{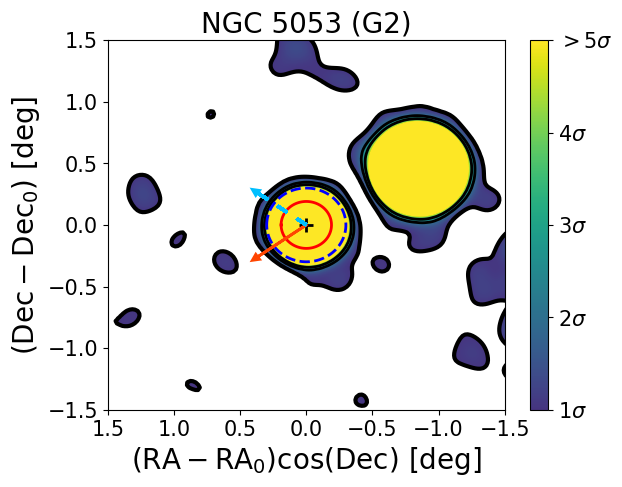}
\includegraphics[width=0.32\textwidth]{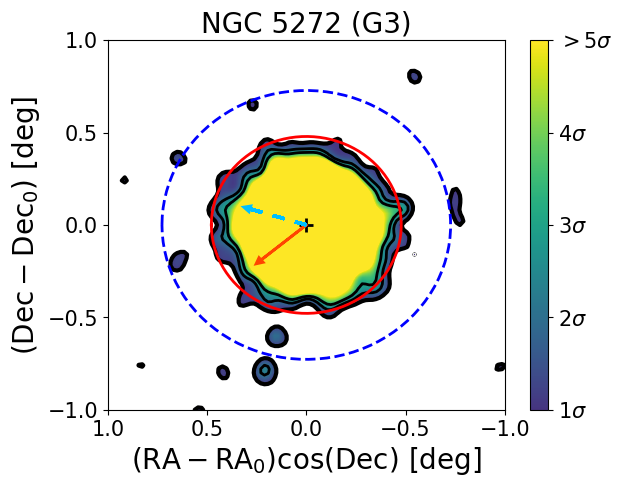}
\includegraphics[width=0.32\textwidth]{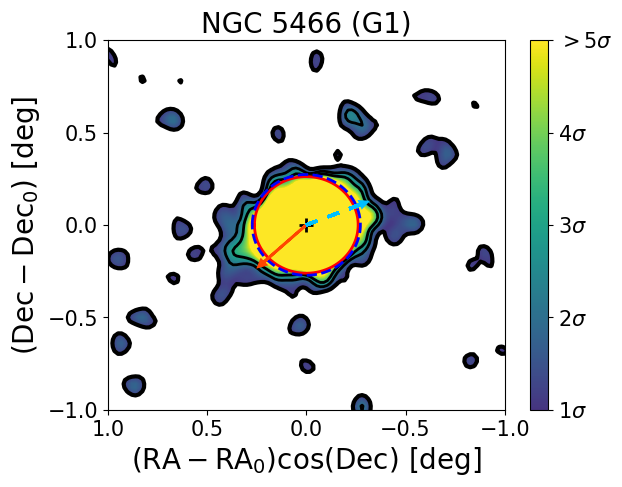}
\includegraphics[width=0.32\textwidth]{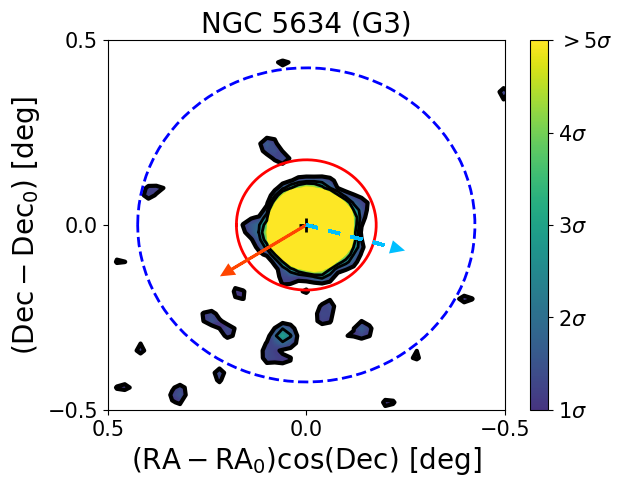}
\includegraphics[width=0.32\textwidth]{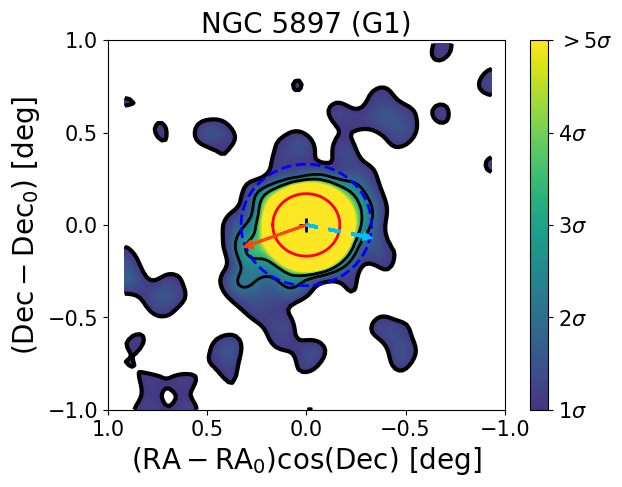}
\includegraphics[width=0.32\textwidth]{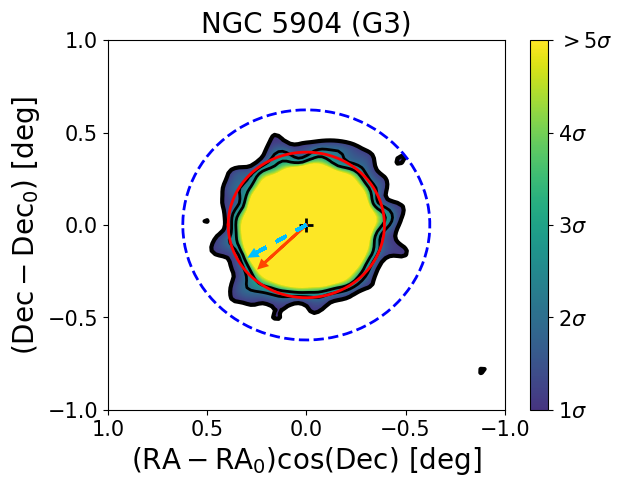}

\center
\caption{Signal-to-noise ratio distributions for the remaining clusters in the sample, complementing those shown in Fig. \ref{figure2}. Among them, a prominent overdensity appears below NGC 362, which is caused by contamination from the Small Magellanic Cloud. In addition, NGC 5024 and NGC 5053 have very similar CMDs and are commonly regarded as an interacting pair of clusters \citep{2010AJ....139..606C}, and thus they appear in each other's field.}
\label{figure2appendix}
\end{figure*}

\begin{figure*}[htbp]
\ContinuedFloat
\center
\includegraphics[width=0.32\textwidth]{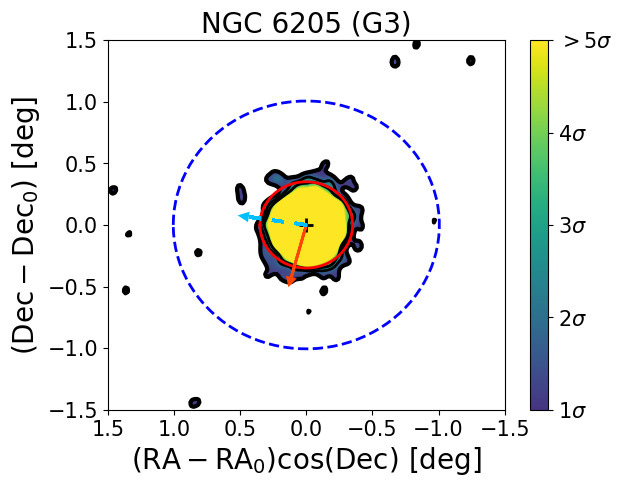}
\includegraphics[width=0.32\textwidth]{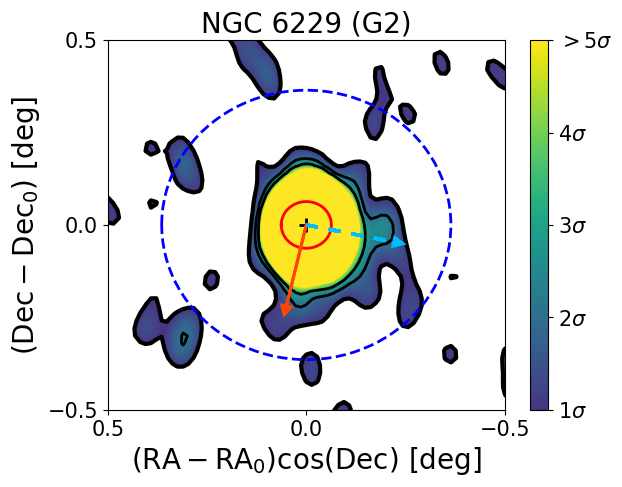}
\includegraphics[width=0.32\textwidth]{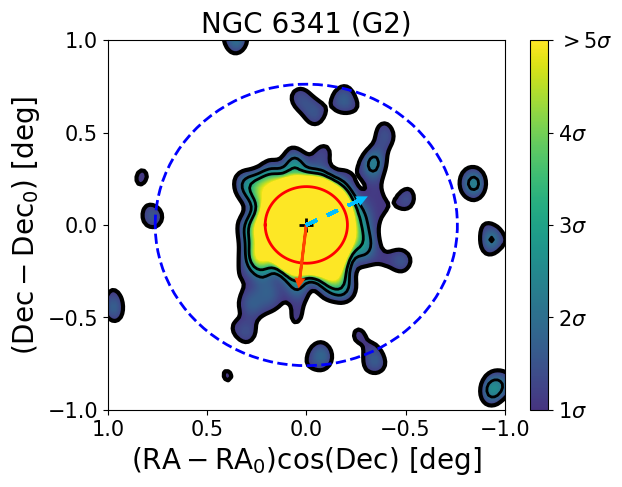}
\includegraphics[width=0.32\textwidth]{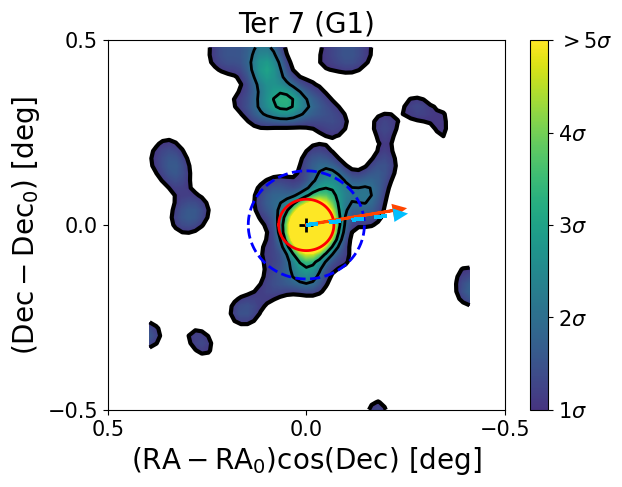}
\includegraphics[width=0.32\textwidth]{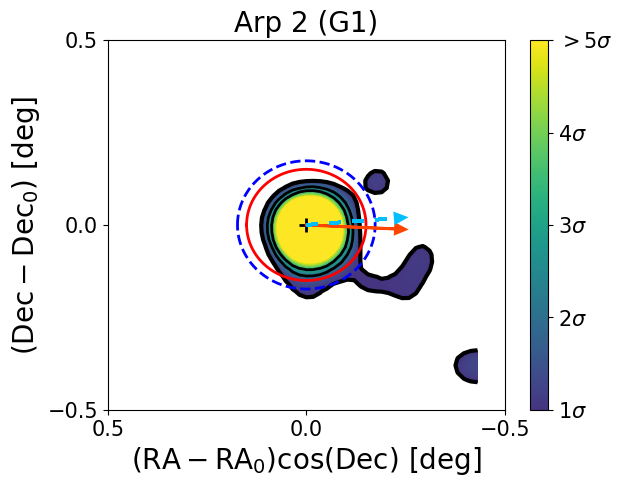}
\includegraphics[width=0.32\textwidth]{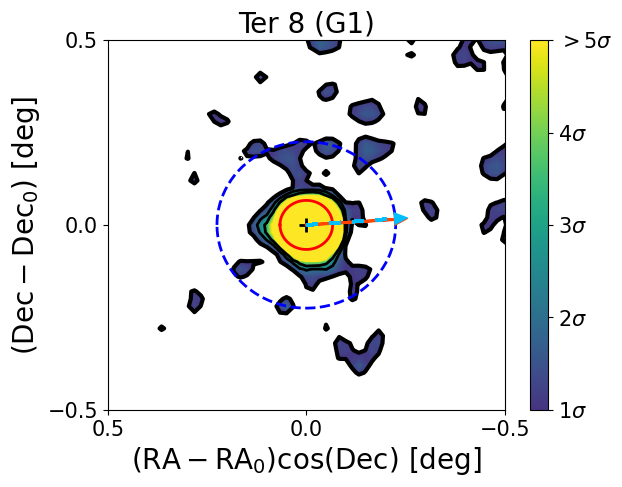}
\includegraphics[width=0.32\textwidth]{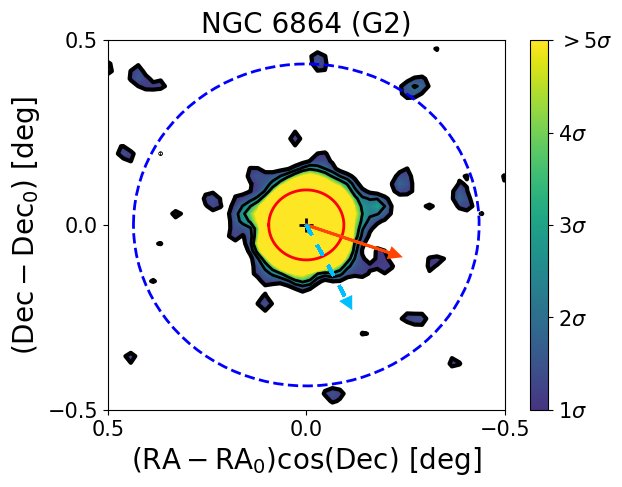}
\includegraphics[width=0.32\textwidth]{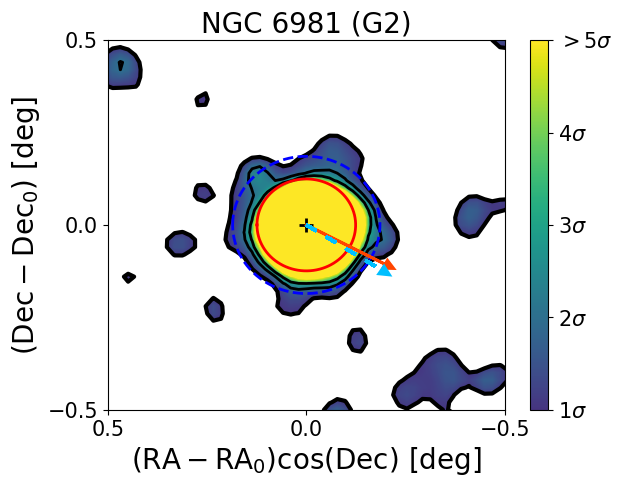}
\includegraphics[width=0.32\textwidth]{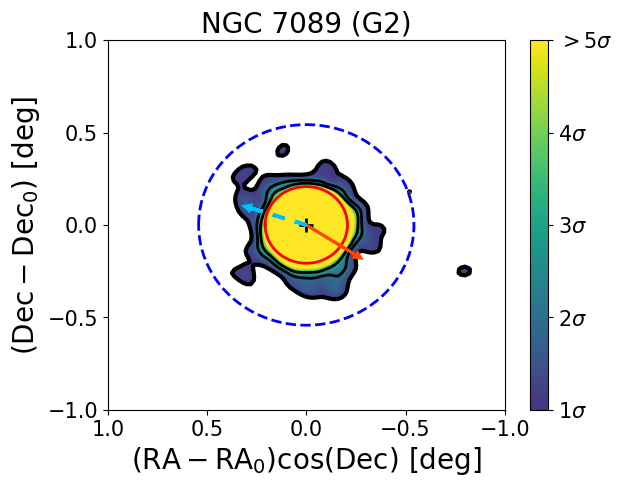}
\includegraphics[width=0.32\textwidth]{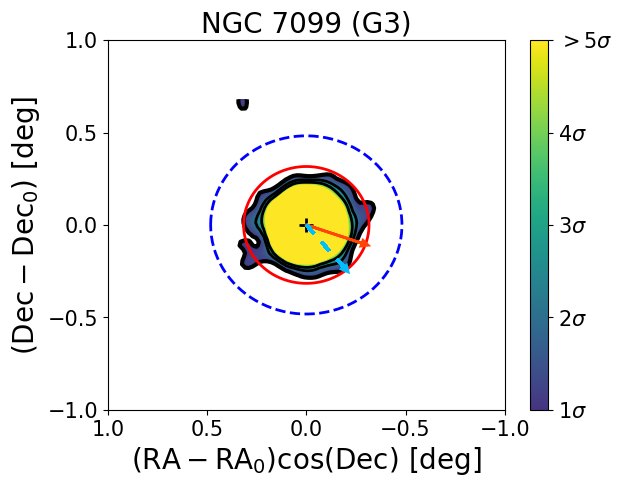}
\includegraphics[width=0.32\textwidth]{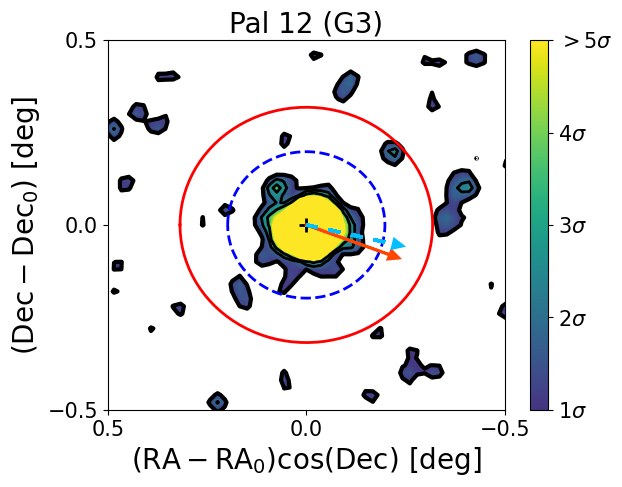}
\includegraphics[width=0.32\textwidth]{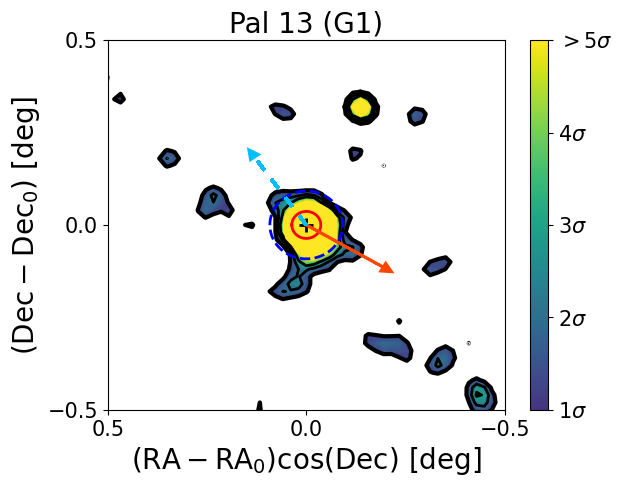}
\includegraphics[width=0.32\textwidth]{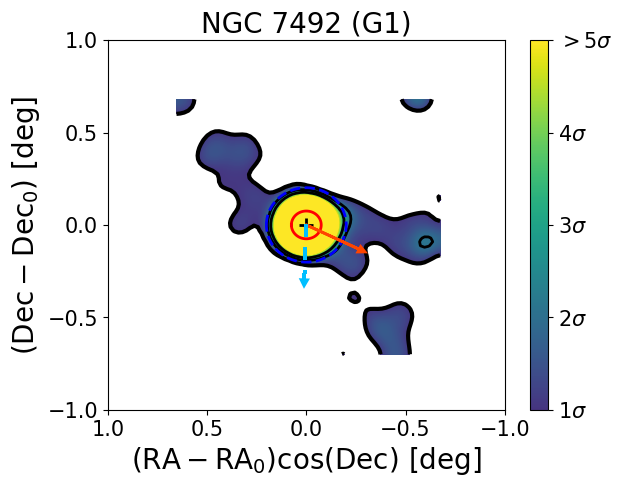}

\center
	\caption{Continued.}
\end{figure*}

\clearpage

\end{appendix}
\end{document}